\documentclass[aps,pre,twocolumn,superscriptaddress,amsmath,amssymb,nofootinbib,noeprint]{revtex4-2}
\usepackage{graphicx,enumitem}
\usepackage{standalone}
\usepackage{multirow,dcolumn}
\usepackage{txfonts}
\usepackage[hyperindex,breaklinks]{hyperref}
\usepackage{soul,cancel}
\usepackage{epsf,amsmath,amsfonts,amssymb,bm,cancel,verbatim,color,multirow,pifont,graphicx,cleveref,tabularx}
\usepackage[dvipsnames]{xcolor}
\usepackage{rotating}
\usepackage{babel}

\begin{document}
\title{Effective epidemic containment strategy in hypergraphs}
	
\author{Bukyoung Jhun}
\email{jhunbk@snu.ac.kr}
\affiliation{CCSS, CTP and Department of Physics and Astronomy, Seoul National University, Seoul 08826, Korea}


\begin{abstract}
Recently, hypergraphs have attracted considerable interest from the research community as a generalization of networks capable of encoding higher-order interactions, which commonly appear in both natural and social systems.
Epidemic dynamics in hypergraphs has been studied by using the simplicial susceptible-infected-susceptible ($s$-SIS) model; however, the efficient immunization strategy for epidemics in hypergraphs is not studied despite the importance of the topic in mathematical epidemiology.
Here, we propose an immunization strategy that immunizes hyperedges with high simultaneous infection probability (SIP).
This strategy can be implemented in general hypergraphs.
We also generalize the edge epidemic importance (EI)-based immunization strategy, which is the state of the art in complex networks. However, it does not perform as well as the SIP-based method in hypergraphs despite its high computational cost.
We also show that immunizing hyperedges with high H-eigenscore effectively contains the epidemics in uniform hypergraphs.
A high SIP of a hyperedge suggests that the hyperedge is a "hotspot" of the epidemic process.
Therefore, SIP can be used as a centrality measure to quantify a hyperedge's influence on higher-order dynamics in general hypergraphs.
The effectiveness of the immunization strategies suggests the necessity of scientific, data-driven, systematic policy-making for epidemic containment.
\end{abstract}

\maketitle

\section{Introduction}

    In the past two decades, extensive research has been devoted to spreading processes in complex networks~\cite{Dorogovtsev2008,Zimmermann2004,Dorogovtsev2002,Klemm2003,Nekovee2007} to model the spread of epidemic diseases~\cite{Pastor-Satorras2015c} and innovations~\cite{Katona2011,Rogers2004}, opinion formation~\cite{Acemoglu2013,Grabowski2006,Watts2007,Chen2020}, and many other physical and social phenomena~\cite{Hoang2003,Newman2003,Boccaletti2006,Velasquez-Rojas2020}.
    Researchers now have access to large-scale datasets of interactions, such as mobility, collaborations, and temporal contacts that were unavailable in the past~\cite{Jia2019,Leskovec2007,Fowler2006}, and complex network representations of interactions enable the researchers to effectively study various dynamical processes.
    The large body of research devoted to spreading processes on complex networks provided quantitative analysis for policy-making especially in the public-health domain.
    Furthermore, the epidemic processes provide deeper understanding of critical phenomena and phase transition behaviors, such as the effect of structural heterogeneity on the transition point~\cite{Pastor-Satorras2001,Moreno2002a} and discontinuous phase transitions induced by cascade dynamics~\cite{Bar2014,Choi2017,Lee2017}.
    
    A hypergraph is a generalization of network that can describe higher-order interactions between more than two agents, which widely appear in both natural and social systems, that networks cannot~\cite{Lambiotte2019,Battiston2020,Bianconi2021,HernandezSerrano2020}. 
    A hypergraph consists of nodes and hyperedges, and a hyperedge of size $d$ connects $d$ nodes simultaneously.
    The hyperedges of a hypergraph can have various sizes, but if all the hyperedges in a hypergraph have the same size $d$, it is called a $d$-uniform hypergraph.
    In a collaboration hypergraph~\cite{Patania2017,Benson2018}, for instance, a hyperedge of size $d$ encodes a $d$-author paper, and the nodes of the hyperedge encodes the authors of the paper.
    Hypergraphs have been used to describe neural and biological interactions~\cite{Petri2014,Klimm2020}, evolutionary dynamics~\cite{Burgio2020,Alvarez-Rodriguez2021}, and other dynamical processes~\cite{Carletti2020,St-Onge2021,DeArruda2020,Lee2021}.
    Recently, the simplicial susceptible-infected-susceptible ($s$-SIS) model~\cite{Iacopini2019} was introduced to describe higher-order epidemic process in hypergraphs.
    The model has attracted extensive interest from the research community due to its simplicity and novel phase transition behavior~\cite{Jhun2019a,Matamalas2020,Landry2020,Chowdhary2021,Wang2021}.
    
    An important topic in epidemiology is immunization, and it has been studied for various epidemic models in complex networks~\cite{Cohen2003,Madar2004,Chen2008,Masuda2009,Pastor-Satorras2002,VanMieghem2011,Matamalas2018,Costa2020,Shim2021}.
    If a node in the network is immunized, the node cannot turn into the infected state, and if an edge is immunized, the infection does not spread through the immunized edge.
    Edge immunization models epidemic containment measures such as travel regulation and social distancing.
    If a node or edge is immunized, it does not only prevent nodes directly connected to them from being infected.
    If a portion of nodes or edges greater than a threshold $p_c$ is immunized, the epidemic state in the network vanishes.
    This effect is called \textit{herd immunity}, and the threshold is called the \textit{herd immunity threshold} (HIT).
    The objective of an efficient immunization strategy is to achieve herd immunity by immunizing a minimal portion of nodes or edges, i.e. minimizing HIT $p_c$.
    Such strategies can be used to vaccinate people with limited resources or prevent a pandemic by minimally regulating air traffic or social gatherings.
    The same theory can be used to promote spreading processes.
    If the spreading process models information flow, for instance, the objective is usually to optimize the spreading of information in a system.
    In such cases, we buttress the nodes or edges targeted by the efficient immunization strategies instead of immunizing them.
    Alternatively, in a reverse point of view, an adversarial attack can be made on such nodes/edges to hamper the information flow in the system.
    However, the efficient immunization strategy for epidemic processes in hypergraph has not been studied, despite the topic's importance in mathematical epidemiology.
    
    Here, we propose an immunization strategy that targets hyperedges with high simultaneous infection probability (SIP), which is the probability that all the nodes in a hyperedge are in the infected state.
    This probability is calculated by the individual-based mean-field (IBMF) theory~\cite{Wang2003,Gomez2010}.
    This strategy can be implemented to contain epidemics of $s$-SIS model in general hypergraphs.
    We also show that immunizing hyperedges with the highest H-eigenscores, which is defined as the product of the elements of the H-eigenvector of the adjacency tensor with the largest H-eigenvalue of all the nodes in the hyperedge, effectively achieves herd immunity in uniform hypergraphs.
    This method generalizes the edge eigenscore in a complex network and can be implemented to contain epidemics in uniform hypergraphs.
    However, this method cannot be implemented in arbitrary hypergraphs with various hyperedge sizes.
    We also generalize the EI-based method~\cite{Matamalas2018}, which is the state-of-the-art immunization strategy for complex networks. However, we find that this method does not perform as efficiently as H-eigenscore and SIP-based strategies for hypergraphs despite its higher computational cost.
    If a hyperedge has a high SIP, it suggests that the hyperedge is a 'hotspot' of the epidemic process.
    Therefore, SIP can be used as a centrality measure to quantify a hyperedge's influence on higher-order dynamics in general hypergraphs.
    The effectiveness of the immunization strategies suggests the necessity of quantitative and systematic policies for epidemic containment measures.
    
    This paper is organized as follows:
    First, we introduce the epidemic model in a hypergraph in Sec.~\ref{sec:def_ssis}.
    Next, we introduce the hypergraph static model in Sec.~\ref{sec:static_model} and the hypergraph popularity-similarity optimization ($h$-PSO) model in Sec.~\ref{sec:hpso}.
    We show that the $h$-PSO model generates a hypergraph with a power-law degree distribution and a tunable clustering coefficient.
    In Sec.~\ref{sec:mean_field_theory}, we extend the individual-based mean-field (IBMF) and pair-based mean-field (PBMF) theories to general hypergraphs, which is required for the immunization strategies.
    The immunization strategies are illustrated in Sec.~\ref{sec:strategy}.
    The strategies' performance in complex networks and hypergraphs are tested in Sec.~\ref{sec:numerical_results}.
    A summary and the final remarks are presented in Sec.~\ref{sec:conclusion}.
	
	\section{Model}
	
    \subsection{Epidemic dynamics in networks and hypergraphs\label{sec:def_ssis}}

    A contagion process through dyadic interaction (represented by a network or a metapopulation model) is called a \textit{simple contagion} process.
    The SIS model is one of the most extensively studied simple contagion model in complex networks along with the susceptible-infected-recovered (SIR) model~\cite{Pastor-Satorras2015c}.
    In the SIS model, each node in the network is in the either susceptible (S) or infected (I) state.
    If a node is infected, it turns into the susceptible state with a constant recovery rate $\mu$.
    If a node is susceptible, it is infected with infection rate $\beta$ from each of its infected neighbors.
    The rates of contagion and infection only depend on the current configuration $\left\{X_1, X_2, \dots, X_N\right\}$, where $X_i$ is the state of node $i$ ($X_i\in\{\mathrm{S},\mathrm{I}\}$), of the epidemic states and not on the past configurations, i.e. they are treated as Poisson processes.
    If the infection rate is higher than a certain value (i.e., epidemic threshold), the system can reach a stationary state, allowing several theoretical approaches~\cite{Pastor-Satorras2001d,Matamalas2018,Valdano2015}.
    
\begin{figure}
		\includegraphics[width=\columnwidth]{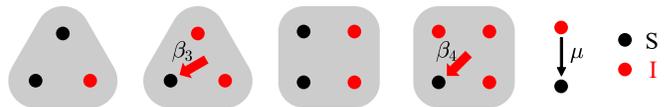}
		\caption{
		Schematic representation of simplicial susceptible-infected-susceptible ($s$-SIS) model. 
        Black dots represent susceptible nodes, red dots represent infected nodes, and the grey area represents hyperedges.
		The contagion along a hyperedge occurs if and only if exactly one node is susceptible and all the other nodes are infected.
        If the condition is met, the susceptible node is turned to the infected state with rate $\beta_d$, where $d$ is the size of the hyperedge.
        If the hyperedge is immunized, the contagion through the hyperedge does not occur.
        Additionally, the recovery process $\rm S\rightarrow I$ is defined identically to the traditional susceptible-infected-susceptible (SIS) model.
		}
		\label{fig:def_ssis}
	\end{figure}

    Many contagion phenomena that cannot be reduced to a simple contagion process have been observed, especially in social systems~\cite{Lionberger1968,Heath2001,MacDonald1964,Crane1999}. 
    More complicated models of contagion, namely \textit{complex contagion} processes including the threshold model and generalized epidemic model have been proposed.
    Among them, the recently introduced $s$-SIS model was introduced as a complex contagion model and has attracted extensive interest due to its simplicity, analytic tractability, and novel critical phenomena~\cite{Iacopini2019,Jhun2019a, Matamalas2020,Landry2020}.
    In the model, the contagion occurs through hyperedges in hypergraphs, which have attracted considerable interest from the research community as a generalization of networks due to their capability of encoding higher-order interactions between more than two agents~\cite{Lambiotte2019,Battiston2020,Bianconi2021,HernandezSerrano2020}.    
    The model is illustrated in Fig.~\ref{fig:def_ssis}.
    In the $s$-SIS model, a node in the hypergraph is in a susceptible or an infected state, as in the traditional SIS model.
    If a susceptible node has a hyperedge of size $d$ where all the other $d-1$ nodes in the hyperedge is in the infected state, the node is changed to the infected state with rate $\beta_d$.
    Here, we study the discrete-time version of the model where time $t$ is integer.
    If a susceptible node at time $t$ has $n$ hyperedges that satisfy the contagion condition, each hyperedge has a probability $\beta_d$ to turn the susceptible node to the infected state at time $t+1$.
    Also, an infected node at time $t$ turns into a susceptible node at time $t+1$ with probability $\mu$.

    \subsection{Hypergraph static model\label{sec:static_model}}

    Many real-world interactions, whether dyadic or high-order, exhibit high heterogeneity characterized by power-law behavior.
	To model such highly heterogeneous hypergraphs, the hypergraph static model~\cite{Jhun2019a} was introduced as a hypergraph model with a degree (number of hyperedges connected to the node) distribution with a power-law tail, namely a scale-free hypergraph.
    It is a generalization of the static model of complex networks~\cite{Goh2001c,Lee2006}, and has been used as a canonical method to generate scale-free networks due to its simplicity and analytical tractability~\cite{Goh2006,Yook2018,Lee2018,Kim2017}.
	In the hypergraph static model, 
    \begin{enumerate}
        \item[(i)] Parameter $p_i$ is assigned to each node in the hypergraph. This parameter controls the nodes' fitness to have a high degree.
        \item[(ii)]  Pick $d$ nodes with probability $p_{i_1}\cdots p_{i_d}$ . If a hyperedge $\{i_1,\cdots,i_d\}$ is not already present in the hypergraph, add it to the hypergraph.
        \item[(iii)] Repeat step (ii) until the number of hyperedges reaches $NK$.
    \end{enumerate}
	If we set  $p_{i}= N i^{-\alpha} / \zeta_{N}(\alpha)  \simeq (1-\alpha)i^{-\alpha}/N^{-\alpha}$, where $\zeta_{N}(\alpha)=\sum_{j=1}^{N}j^{-\alpha}$ and $0<\alpha<1$ (hence, $\sum p_i = 1$ and $0<p_i<1$), we obtain a hypergraph with power-law degree distribution.
	Because the probability of each node being chosen in step (ii) is independent and identically distributed (i.i.d), node $i$ is chosen with probability $p_i$ in each iteration, hence each node $i$ has expected degree $\left<k_i\right>$, and the distribution of the expected degree $P_d(\left<k_i\right>) \sim \left<k_i\right>^{-\gamma}$ with $\gamma=1+1/\alpha$. 
	The minimum degree $k_m = {N^{1-\alpha}\langle k \rangle }/{\sum_{j=1}^{N}j^{-\alpha}}$ converges to a finite value $\frac{\gamma-2}{\gamma-1}\langle k \rangle$, and the maximum degree $k_{\rm max}={N\langle k \rangle}/{\sum_{j=1}^{N}j^{-\alpha}}$ diverges in the thermodynamic limit $N\rightarrow\infty$.
	Thus, we obtain a scale-free network with mean degree $\left< k \right> = dK$ and degree exponent $\gamma$.
	We obtain an Erd{\H o}s-R\'enyi-type random hypergraph in the $\gamma\rightarrow\infty$ limit.

   \subsection{Hypergraph popularity-similarity optimization ($h$-PSO) model\label{sec:hpso}}

    \begin{figure}
		\includegraphics[width=\columnwidth]{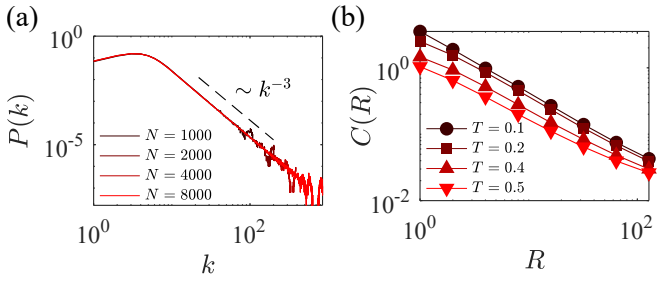}
		\caption{
		(a) The degree distribution of the 3-uniform hypergraph popularity-similarity optimization ($h$-PSO) model.
        The mean degree $\left<k\right>=6$, temperature $T=0.5$, and the parameter $\gamma=3$.
        The tail of the distribution follows a power law with an exponent of 3.
        (b) The clustering coefficient of 3-uniform $h$-PSO as a function of the scale parameter $R$.
        The number of nodes $N=2000$, mean degree $\left<k\right>=6$, and the temperature $T=0.5$.
		}
		\label{fig:degree_distribution}
	\end{figure}
    
	In addition to a highly heterogeneous degree distribution, agents in many real-world systems have a higher chance of being connected if they are similar.
	The similarity of two nodes is characterized by their closeness in their latent coordinates.
	The objective of graph node embedding algorithms~\cite{Papadopoulos2015,Perozzi2014,Grover2016} is to discover the latent coordinates of a network.
    For instance, hub airports are connected to a disproportionately large number of airports around the world (heterogeneous degree distribution), but two small airports can be connected by an airline if they are geographically close.
    Also, two researchers who are not particularly prolific can coauthor a paper if they are close. 
    This effect is called \textit{homophily} and results in non-vanishing clustering coefficients in both networks and hypergraphs.
    To account for such phenomena, a hypergraph model with a scale-free degree distribution and tunable non-vanishing clustering coefficient needs to be introduced.
    Furthermore, the immunization strategies need to be tested in clustered hypergraphs because it is known that epidemic dynamics and the performance of immunization strategies differ in clustered and unclustered networks~\cite{Matamalas2018}.
    
    The clustering coefficient $C(H)$ of a hypergraph $H$ is defined as follows~\cite{Estrada2005}:
    \begin{equation}
        C(H) = \frac{3 \times \operatorname{number~of~hypertriangles}}{\operatorname{number~of~undirected~2-paths}}\,,
    \end{equation}
    where a hypertriangle is a set of three distinct nodes $v_1$, $v_2$, $v_3$ and three distinct hyperedges $E_{12}$, $E_{23}$, $E_{31}$ that satisfies $v_1, v_2 \in E_{12}$, $v_2, v_3 \in E_{23}$, and $v_3, v_1 \in E_{31}$.
    A undirected 2-path is a set of three distinct nodes $v_1$, $v_2$, $v_3$ and two distinct hyperedges $E_{12}$, $E_{23}$ that satisfies $v_1, v_2 \in E_{12}$ and $v_2, v_3 \in E_{23}$.
    The clustering coefficient can be greater than 1 in hypergraphs because a undirected 2-path can have multiple closures.
    If there are only size-2 hyperedges in the hypergraph (i.e., if the hypergraph is a network), $C_2$ becomes the transitivity coefficient~\cite{Harary1979}, which is widely used in social network analysis.
    Note that there is another definition of the clustering coefficient $C_d^{(i)}$ that generalizes the local clustering coefficient of graphs~\cite{Burgio2020}.
    
   To generate a scale-free hypergraph with a non-vanishing clustering coefficient, we introduce the hypergraph popularity-similarity optimization model ($h$-PSO), which is a hypergraph version of the popularity-similarity optimization (PSO) model in complex networks~\cite{Papadopoulos2015,Papadopoulos2012}.
   The $d$-uniform $h$-PSO model is generated as follows:
    \begin{enumerate}
        \item[(i)] Popularity parameter $p_i$ is assigned to each node in the hypergraph. If a node has a high $p_i$, the node tends to have a high degree.
        \item[(ii)] Latent coordinate $\textbf{x}_i$ is assigned to each node in the hypergraph. If two nodes $i$ and $j$ are close in the latent coordinate (i.e., $\left| \textbf{x}_j - \textbf{x}_i \right|$ is small) two nodes will likely be connected by hyperedges.
        \item[(iii)]  Pick a node $i$ with probability $p_{i}$.
        \item[(iv)]  Pick $d-1$ nodes $j_1,\cdots,j_{d-1}$, each with probability $\left[1+\left(|\textbf{x}_{j_\ell}-\textbf{x}_i| / R p_{i}p_{j_\ell}\right)^{1/T}\right]^{-1}$ . If a hyperedge $\{i,j_1\cdots,j_{d-1}\}$ is not already present in the hypergraph, add it to the hypergraph.
        \item[(v)] Repeat steps (iii)--(iv) until the number of hyperedges reaches $NK$.
    \end{enumerate}
	Here, we choose the latent coordinates on a ring; the latent coordinates are randomly chosen without replacement from $\theta\in\{1,2,\cdots,N \}$, and the distance between two nodes $i$ and $j$ is defined as $\min\left( \left|\theta_j - \theta_i\right| , N-\left|\theta_j - \theta_i\right| \right)$.
    If we set $p_i = N i^{-\alpha} / \zeta_{N}(\alpha) \simeq (1-\alpha)i^{-\alpha}/N^{-\alpha}$, the resulting hypergraph is a scale-free hypergraph with degree exponent $\gamma=1+1/\alpha$.
    The clustering coefficient can be controlled by the scale parameter $R$ and the temperature $T$; if $R$ and $T$ are large, the clustering coefficient is small.
    The degree distribution and the clustering coefficient of the $h$-PSO model with hyperedge size 3 are illustrated in Fig.~\ref{fig:degree_distribution}.
    The degree distribution has a power-law tail with exponent $\gamma$, and the clustering coefficient can be controlled by adjusting $R$.

	\section{Individual- and pair-based mean-field theories\label{sec:mean_field_theory}}
	
	In this section, we explain the individual-based mean-field (IBMF) theory and pair-based mean-field~\cite{Matamalas2018} (PBMF) theory for hypergraphs, which are used in immunization strategies.
    IBMF tracks the probability of infection $p_i$ of each node in the network.
	By ignoring the statistical correlation of the probability between two nodes [$P(X_i,X_j) = P(X_i)P(X_j)$, where $X_i,X_j\in\{S,I\}$], the IBMF equation for the SIS model can be expressed as
	\begin{equation}
	p_i (t+1) = [1-p_i(t)] \left[1 - \prod_{j\in\mathcal{N}(i)} \left(1 - \beta p_j(t)\right)\right] + (1-\mu) p_i(t) \label{eq:IBMF}\,,
	\end{equation}
	where $\mathcal{N}(i)$ is the set of nodes connected to node $i$ (nearest-neighbors of $i$).
    For continuous phase transitions, where $p_i$ vanishes in the vicinity of the phase transition, the equation can be linearized as $ p_i (t+1) = \sum_j \left(\beta a_{ij} + (1-\mu) \delta_{ij}\right) p_j$ and the epidemic threshold $\frac{\beta}{\mu}$ is the inverse of the largest eigenvalue of the adjacency matrix $a_{ij}$.
    Because IBMF ignores the positive correlations of the state (neighbors of infected node have greater chance of being in the infected state) in the actual system, it tends to overestimate the density of infection.
	The theory can be straightforwardly extended to the $s$-SIS model:
	\begin{align}
    	p_i (t+1) &= [1-p_i(t)] \prod_{\left\{ j_{1},\cdots,j_{d-1}\right\} \in\bar{\mathcal{N}}(i)}\left(1-\beta_{d} p_{j_1}(t)\cdots p_{j_{d-1}}(t) \right) \nonumber \\
    	&+ (1-\mu) p_i(t) \,,
    \end{align}
    where $\bar{\mathcal{N}}(i)$ is the set of 'hyperneighbors' of $i$; if a hyperedge $\{i,j_1,\cdots,j_{d-1}\}$ is in the hypergraph, $\{j_1,\cdots,j_{d-1}\}\in\bar{\mathcal{N}}(i)$.
	IBMF is often employed to describe the dynamics and phase transitions in classical stochastic processes~\cite{Wang2003,Gomez2010}, as well as driven-dissipative quantum dynamics~\cite{Perez-Espigares2017a}.
	The method predicts the properties of the epidemic states more accurately than homogeneous mean-field theory or degree-based mean-field theory~\cite{Pastor-Satorras2001}, which is often referred to as heterogeneous mean-field theory.
		
	PBMF, often referred to as an epidemic-link equation, is known to predict the properties of the epidemic states more precisely than IBMF.
	In PBMF, we track the probability of the infection $p_i(t)$ of each node the same as for IBMF, and for pairs of nodes $(i,j)$ that are connected in the network we set the differential equations for the probability that both of the nodes are infected as $\psi_{ij}(t)=P(X_i=I,X_j=I)$.
	Probabilities for other cases for a node pair $P(X_i=S,X_j=S)$, $P(X_i=S,X_j=I)$, and $P(X_i=I,X_j=S)$ can be expressed in terms of the $p_i$ and $\psi_{ij}$:
	\begin{align}
	    P(X_i=S,X_j=S) &= 1-p_i(t)-p_j(t)+\psi_{ij}(t) \,, \\
	    P(X_i=S,X_j=I) &= p_j(t) - \psi_{ij}(t) \,, \\
	    P(X_i=I,X_j=S) &= p_i(t) - \psi_{ij}(t) \,.
	\end{align}
	This method exploits the sparsity of the network (the number of variables and the equations in this method is proportional to the number of the nodes in the system); hence, it is scalable to large networks.

	Then, the equations for the nodes are expressed as
    \begin{equation}
      p_{i}(t+1)=\left(1-q_{i}(t)\right)(1-p_{i}(t))+(1-\mu)p_{i}(t) \,,
    \end{equation}
    and the equations for the pairs are expressed as
    \begin{widetext}
    \begin{align}
       \psi_{ij}(t+1) &= \left(1-q_{ij}(t)\right)\left(1-q_{ji}(t)\right)\left(1-p_{i}(t)-p_{j}(t)+\psi_{ij}(t)\right)+\left(1-(1-\beta)q_{ij}(t)\right)(1-\mu)\left(p_{j}(t)-\psi_{ij}(t)\right)\\
       &+\left(1-(1-\beta)q_{ji}(t)\right)(1-\mu)\left(p_{i}(t)-\psi_{ij}(t)\right)+(1-\mu)^{2}\psi_{ij}(t)\,,
    \end{align}
    \end{widetext}
    where
    \begin{align}
        q_{i}(t)&=\prod_{j\in\mathcal{N}(i)}\left(1-\beta\frac{p_{j}(t)-\psi_{ij}(t)}{1-p_{i}(t)}\right) \,,\\
        q_{ij}(t)&=\prod_{\substack{r\in\mathcal{N}(i)\\r\neq j}}\left(1-\beta\frac{p_{j}(t)-\psi_{ij}(t)}{1-p_{i}(t)}\right) \,.
    \end{align}
    $q_i(t)$ is the probability that node $i$ is not infected during time step $t\rightarrow t+1$ given that the node $i$ is not infected at time $t$; $q_{ij}(t)$ is the probability that the node $i$ is not infected by a neighbor other than $j$ during the time step $t\rightarrow t+1$ given that the node $i$ is not infected at time $t$.

	Stationary states of the $s$-SIS model have been studied using both IBMF and PBMF in hypergraphs with hyperedges with sizes less than or equal to 3~\cite{Matamalas2020}.
	Implementing the PBMF on general hypergraphs with arbitrary hyperedge sizes, the equations for the nodes are, again,
	\begin{align}
	    p_{i}(t+1)=\left(1-q_{i}(t)\right)(1-p_{i}(t))+(1-\mu)p_{i}(t) \,,
	\end{align}
    and the equations for the pairs that are connected by hyperedges are
    \begin{widetext}
    \begin{align}
    \psi_{ij}(t)&=\left(1-q_{ij}(t)\right)\left(1-q_{ji}(t)\right)\left(1-p_{i}(t)-p_{j}(t)+\psi_{ij}(t)\right) + \left(1-q_{ij}(t) u_{ij}(t)\right)(1-\mu)\left(p_{j}(t)-\psi_{ij}(t)\right) \\
    &+ \left(1-q_{ji}(t)u_{ji}(t)\right)(1-\mu)\left(p_{i}(t)-\psi_{ij}(t)\right) + (1-\mu)^{2}\psi_{ij}(t) \,,
    \end{align}
    where
    \begin{equation}
        q_{i}(t)=\prod_{\left\{ r_{1},\cdots,r_{d-1}\right\} \in\bar{\mathcal{N}}(i)}\left(1-\beta_{d}\frac{P_{ir_{1}\cdots r_{d-1}}^{SI\cdots I}(t)}{P_{i}^{S}(t)}\right)=\prod_{\left\{ r_{1},\cdots,r_{d-1}\right\} \in\bar{\mathcal{N}}(i)}\left(1-\beta_{d}\frac{\prod_{\ell=1}^{d-1}\left(p_{r_{\ell}}(t)-\psi_{ir_{\ell}}(t)\right)\prod_{\ell\neq m}\psi_{r_{\ell}r_{m}}(t)}{\left(1-p_{i}(t)\right)^{d-1}\left(\prod_{\ell=1}^{d-1}p_{r_{\ell}}(t)\right)^{d-2}}\right) \,,
    \end{equation}
    \begin{equation}
        q_{ij}(t)=\prod_{\substack{\left\{ r_{1},\cdots,r_{d-1}\right\} \in\bar{\mathcal{N}}(i)\\r_{1},\cdots,r_{d-1}\neq j}}\left(1-\beta_{d}\frac{P_{ir_{1}\cdots r_{d-1}}^{SI\cdots I}}{P_{i}^{S}}\right)=\prod_{\substack{\left\{ r_{1},\cdots,r_{d-1}\right\} \in\bar{\mathcal{N}}(i)\\r_{1},\cdots,r_{d-1}\neq j}}\left(1-\beta_{d}\frac{\prod_{\ell=1}^{d-1}\left(p_{r_{\ell}}(t)-\psi_{ir_{\ell}}(t)\right)\prod_{\ell\neq m}\psi_{r_{\ell}r_{m}}(t)}{\left(1-p_{i}(t)\right)^{d-1}\left(\prod_{\ell=1}^{d-2}p_{r_{\ell}}(t)\right)^{d-2}}\right) \,,
    \end{equation}
    \begin{equation}
        u_{ij}(t)=\prod_{\left\{ j,r_{1},\cdots,r_{d-2}\right\} \in\bar{\mathcal{N}}(i)}\left(1-\beta_{d}\frac{P_{ijr_{1}\cdots r_{d-2}}^{SI\cdots I}(t)}{P_{ij}^{SI}(t)}\right)=\prod_{\left\{ j,r_{1},\cdots,r_{d-2}\right\} \in\bar{\mathcal{N}}(i)}\left(1-\beta_{d}\frac{\prod_{\ell=1}^{d-2}\left(p_{r_{\ell}}(t)-\psi_{ir_{\ell}}(t)\right)\prod_{\ell=1}^{d-2}\psi_{jr_{\ell}}(t)\prod_{\ell\neq m}\psi_{r_{\ell}r_{m}}(t)}{\left(1-p_{j}(t)\right)^{d-1}p_{j}(t)^{d-2}\left(\prod_{\ell=1}^{d-2}p_{r_{\ell}}(t)\right)^{d-2}}\right) \,,
    \end{equation}
    \end{widetext}
    where $P_{r_{1}\cdots r_{d}}^{X_{1}\cdots X_{d}}$ is the probability that nodes $r_{1}\cdots r_{d}$ are each in state $X_{1}\cdots X_{d}$.
    $q_i(t)$ represents the same probability in the network PBMF, $q_{ij}(t)$ is the probability that the node $i$ is not infected by hyperedges that do not contain node $j$ during time step $t\rightarrow t+1$ given that node $i$ is not infected at time $t$, and $u_{ij}(t)$ is the probability that node $i$ is not infected by hyperedges that contain node $j$ during time step $t\rightarrow t+1$ given that node $i$ is not infected at time $t$.
    We have used the following equation for closure:
    \begin{align}
        P_{r_{1}\cdots r_{d}}^{X_{1}\cdots X_{d}} = \frac{\prod_{\ell\neq m}P(X_{r_{\ell}},X_{r_{m}})}{\left(\prod_{\ell=1}^{d}P(X_{r_{\ell}})\right)^{d-2}} \,.
    \end{align}
    For $d\leq 3$, we recover the identity in Ref.~\cite{Matamalas2020}.
 	
	\section{Immunization strategies\label{sec:strategy}}
	
    An immunization strategy is defined as a specific rule that determines a set of nodes or edges that will be immunized to eliminate the epidemic from the network.
    Immunized nodes cannot be infected and the infection cannot spread along the immunized edges.
    The immunization of nodes/edges does not only protect the nodes directly connected to them.
    When a sufficiently large fraction $p>p_c$ of the nodes/edges are immune, the system cannot maintain the epidemic state with a non-vanishing density of infection.
    This effect is called \textit{herd immunity}.
    The objective of an immunization strategy is to find an algorithm that minimizes $p_c$.
    Efficient immunization strategy that can be implemented in complex network has been extensively studied for both SIS~\cite{VanMieghem2011,Matamalas2018,Pastor-Satorras2002,Masuda2009} and SIR~\cite{Chen2008,Madar2004,Cohen2003} model.
    However, efficient immunization strategy for epidemics in hypergraphs has not been studied, despite the importance of the subject.
    Here, we develop a simultaneous infection probability (SIP)-based immunization strategy that can be used to efficiently eliminate epidemic states by immunizing edges in networks or hyperedges in hypergraphs.
    The strategy immunizes the edges or hyperedges in the descending order of the SIP, which is the probability that all the nodes in the edge/hyperedge are infected at the same time.
    The probability is calculated by IBMF in both networks and hypergraphs.
    In networks, the strategy is as efficient as the EI-based strategy~\cite{Matamalas2018}, which is the state-of-the-art immunization strategy, while incurring a lower computational cost.
    This method can be implemented in general nonuniform hypergraphs.
    We compare the efficiency of the strategy with several other methods in networks, uniform hypergraphs, and nonuniform hypergraphs.
    However, only the proposed SIP-based strategy can be efficiently implemented in general nonuniform hypergraphs.
 
    The EI of an edge is defined as $I_{ij} = g_{ij} + g_{ji}$, where
    \begin{equation}
        g_{ij}=\beta P\left(X_{i}=S,X_{j}=I\right)\sum_{r\in\mathcal{N}(i)}\beta P\left(X_{r}=S|X_{i}=I\right) \,.
    \end{equation}
    The probabilities are calculated by means of PBMF; $\beta P\left(X_{i}=S,X_{j}=I\right)$ is the probability that the infection spreads from $j$ to $i$ along the edge $(i,j)$, and $\sum_{r\in\mathcal{N}(i)}\beta P\left(X_{r}=S|X_{i}=I\right)$ quantifies the impact of such an event.
    For the $s$-SIS model in hypergraphs, the epidemic importance  is expressed as
    \begin{equation}
        I_{\{i_{1},\cdots,i_{d}\}}=\sum_{\sigma\in S(\{i_{1},\cdots,i_{d}\})}g_{\sigma} \,,
    \end{equation}
    where $S(\{i_{1},\cdots,i_{d}\})$ is the set of all the permutations of the set $\{i_{1},\cdots,i_{d}\}$, and
    \begin{widetext}
    \begin{equation}
        g_{i_{1}\cdots i_{d}}=\beta_{d}P\left(X_{i_{1}}=S,X_{i_{2}}\cdots X_{i_{d}}=I\right)\sum_{\{j_{1}\cdots j_{d^{\prime}-1}\}\in\mathcal{N}(i_{1})}\beta_{d^{\prime}}\sum_{\ell=1}^{d^{\prime}-1}P\left(X_{j_{1}}=I,\cdots,X_{j_{\ell}-1}=I,X_{j_{\ell}}=S,X_{j_{\ell}+1}=I,\cdots,X_{j_{d^{\prime}-1}}=I|X_{i}=I\right) \,.
    \end{equation}
    \end{widetext}
    It was shown that immunizing edges with high EI efficiently eliminates epidemic states in various synthetic and empirical networks.
    Because we use PBMF, as the size of hyperedge $d$ increases, the number of pairs whose probability should be tracked by $\psi_{ij}$ rapidly increases, and the computational cost of the method diverges.

    \begin{figure*}
		\includegraphics[width=\textwidth]{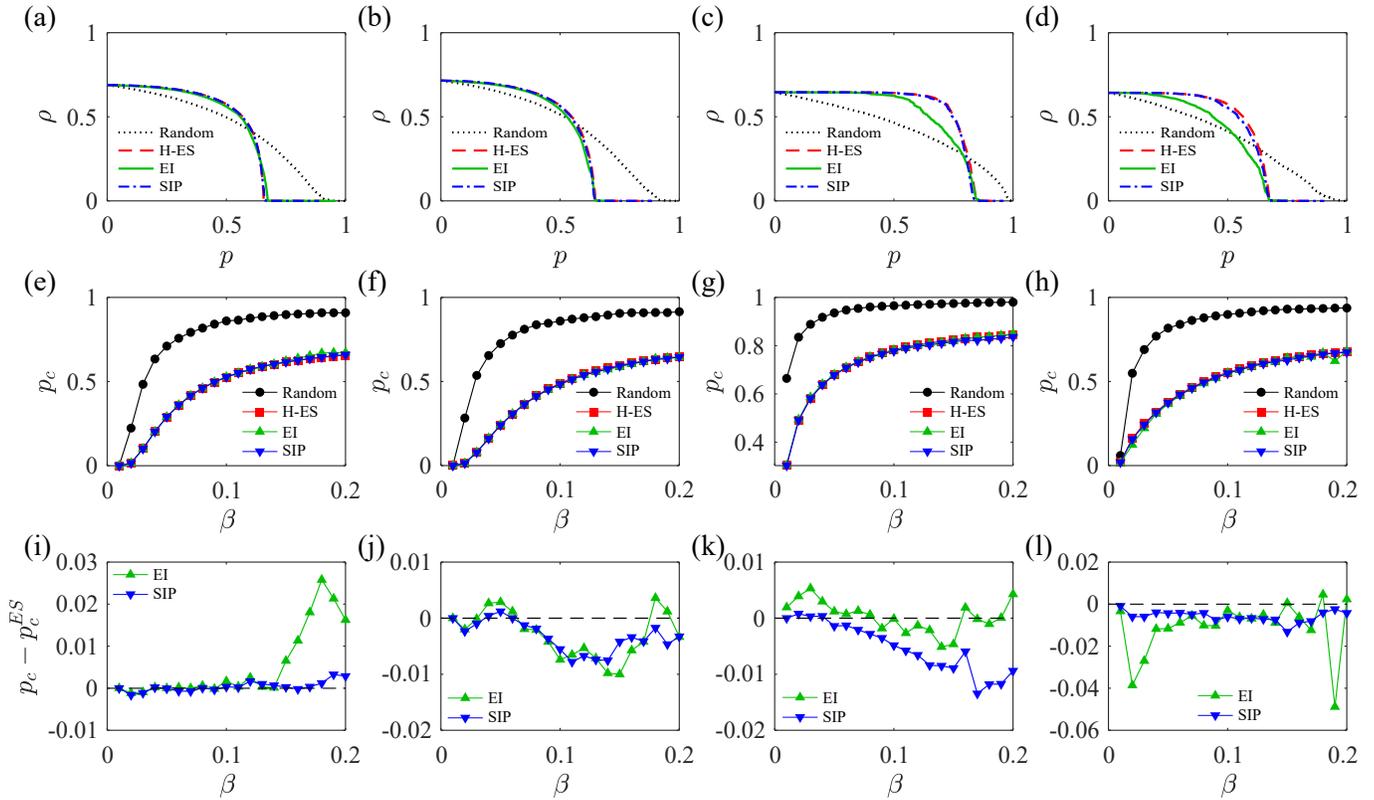}
		\caption{
        Random edge immunization (Random), H-eigenscore (H-ES), EI, and SIP-based strategies tested in various synthetic and empirical networks: (a, e, i) the static model, (b, f, j) clustered power-law network, (c, g, k) an airline network, and (d, h, l) the general relativity collaboration network.
        Because this is a network (i.e. hypergraph with only size-two hyperedges) the H-eigenscore is identical to the usual eigenscore.
		(a, b, c, d) The density of infection $\rho$ versus the removed portion of the edges $p$.
        The recovery rate $\mu=0.2$ and the contagion rate $\beta = 0.2$.
        Efficient immunization strategies usually result in a higher density of infection compared to random immunization for small $p$ but eliminate epidemics with smaller $p_c$.
        (e, f, g, h) HIT $p_c$, which is the minimally required portion of edges to eliminate the epidemics.
        The efficient strategies (i.e. H-eigenscore, EI, and SIP-based strategies) exhibit marginal differences in their HITs.
        To compare the HITs of efficient strategies more thoroughly, we plot the differences between the $p_c$s of EI and SIP-based strategies from the $p_c$ of the H-eigenscore-based strategy in (i, j, k, l).
        The SIP-based strategy is often more efficient than EI-based strategy, despite its lower computational cost.
		}
		\label{fig:plot_network}
	\end{figure*}

    The eigenscore~\cite{VanMieghem2011}, which is widely used as a centrality measure, of a node $i$ is the element of the largest eigenvector $e_i$ of the adjacency matrix, and the eigenscore of an edge $(i,j)$ is the product of the eigenscores of the two nodes of the edge $e_i e_j$.
    By immunizing the edges with the highest eigenscore, the spectral radius of the network is effectively reduced, and the epidemics in the network can efficiently be contained.
    The eigenscore-based strategy can be generalized for implementation in hypergraphs; however, there are multiple types of eigenvectors and eigenvalues in a uniform hypergraph.
    We find that the H-eigenvector is more suitable than the Z-eigenvector~\cite{Lim2005,Qi2005} for $s$-SIS dynamics.
	The H-eigenvector $e_i$ of a $d$-uniform hypergraph is defined as a vector that satisfies
    \begin{equation}
        \left(\boldsymbol{a}\boldsymbol{e}^{d-1}\right)_{i_1} \coloneqq \sum_{i_{2},\cdots,i_{2}=1}^{n}a_{i_{1}i_{2}\cdots i_{d}}e_{i_{2}}\cdots e_{i_{d}}=\lambda e_{i_{1}}^{d-1} \,,
    \end{equation}
    where $\boldsymbol{a}$ is the hypergraph adjacency tensor.
    We define the H-eigenscore of the hyperedge $\{i_1,\cdots,i_d\}$ as the product of the elements of the H-eigenvector with the largest H-eigenvalue: $e_{i_1}\cdots e_{i_d}$.
    For networks where $d=2$, the H-eigenscore becomes the traditional eigenscore.
    Because the adjacency tensor is symmetric and hence diagonalizable~\cite{Comon2008}, the H-eigenvector with the largest H-eigenvalue can be computed by an iterative power method:
    \begin{align}
        \tilde{e}_{i_{1}}^{(m+1)} &= \left(\sum_{i_{2},\cdots,i_{2}=1}^{n}a_{i_{1}i_{2}\cdots i_{d}}e_{i_{2}}^{(m)}\cdots e_{i_{d}}^{(m)}\right)^{\frac{1}{d-1}} \,, \\
        e_i^{(m+1)} &= \frac{\tilde{e}_i^{(m+1)}}{\sqrt{\sum_{j=1}^n \left|\tilde{e}_j^{(m+1)}\right|^2}} \,.
    \end{align}
    Then, $\textbf{e}^{(m)}$ converges to the H-eigenvector with the largest H-eigenvalue as $m\rightarrow \infty$.
    We show that removing high H-eigenscore hyperedges leads to effective epidemic containment in uniform hypergraphs.
    However, for nonuniform hypergraphs, the adjacency tensor is not defined, and the method cannot be implemented in general nonuniform hypergraphs.
	
    We introduce SIP as a measure of a hyperedge's contribution to the continuation of epidemics in the hypergraph.
    The SIP of a size-$d$ hyperedge $\{r_1, \dots, r_d\}$ is the probability that all nodes in the hyperedge are infected, which is calculated by the IBMF
    \begin{equation}
        P_{r_{1}\cdots r_{d}}^{\mathrm{I}\cdots \mathrm{I}} \simeq P^{\mathrm{I}}_{r_1} \cdots P^{\mathrm{I}}_{r_d}\,.
    \end{equation}
    Each infection probability $P^{\mathrm{I}}_{r_\ell}$ can be numerically calculated by solving Eq.~\eqref{eq:IBMF} for its fixed point. 
    Because this method uses IBMF, it incurs less computational cost than the EI-based strategy.
    This measure can be calculated in arbitrary nonuniform hypergraphs whose hyperedges have various sizes. 
    We test the strategies in Sec.~\ref{sec:numerical_results}.

    Other centrality measures have been tested for immunization strategies; however, they were found to be inefficient.
    Immunizing high edge-betweenness edges is ineffective, sometimes less efficient than randomly immunizing edges~\cite{Matamalas2018}.
    The node-infectivity-based method has been tested as well, but it is not as efficient as the eigenscore or EI-based methods.
    
    \section{Numerical Results\label{sec:numerical_results}}

    \begin{figure}
		\includegraphics[width=\columnwidth]{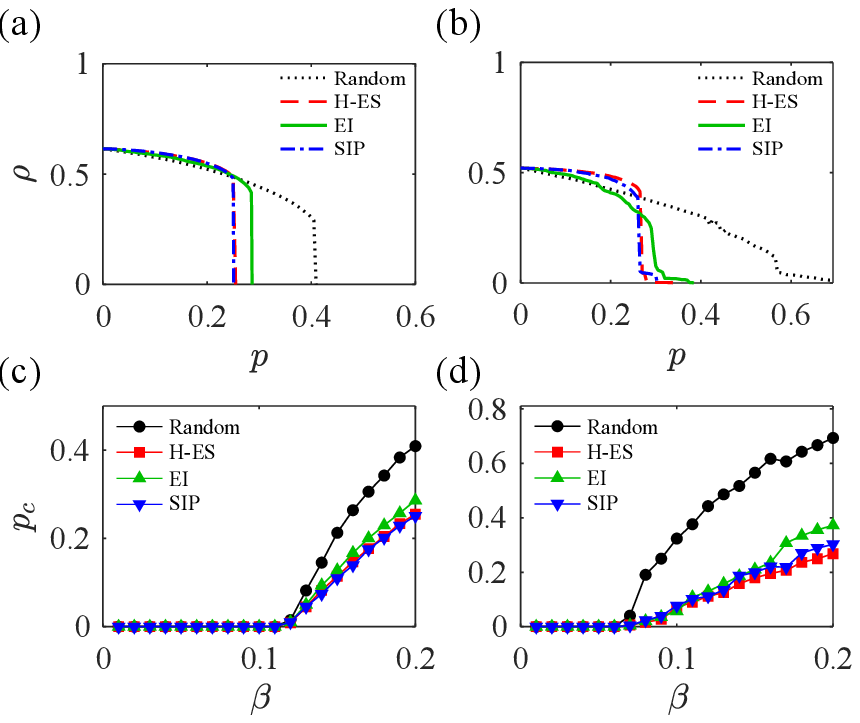}
		\caption{
        Random hyperedge immunization (Random), H-eigenscore (H-ES), EI, and SIP-based strategies tested in 3-uniform hypergraphs: (a, c) the hypergraph static model and (b, d) the hypergraph popularity-similarity optimization ($h$-PSO) model.
		(a, b) The density of infection $\rho$ versus the removed portion of the edges $p$.
        The recovery rate $\mu=0.2$ and the contagion rate $\beta = \beta_3 = 0.2$.
        The efficient strategies generally exhibit a higher density of infection for small $p$, but herd immunity is achieved at lower $p_c$, which is the minimally required portion of hyperedges that needs to be immunized to eliminate epidemics.
        (c, d) HIT $p_c$ as a function of contagion rate $\beta = \beta_3$.
        The H-ES and SIP-based strategies outperform the EI-based strategy, despite their lower computational cost.
		}
		\label{fig:plot_uniform_hypergraph}
	\end{figure}
    
     \begin{figure}
		\includegraphics[width=\columnwidth]{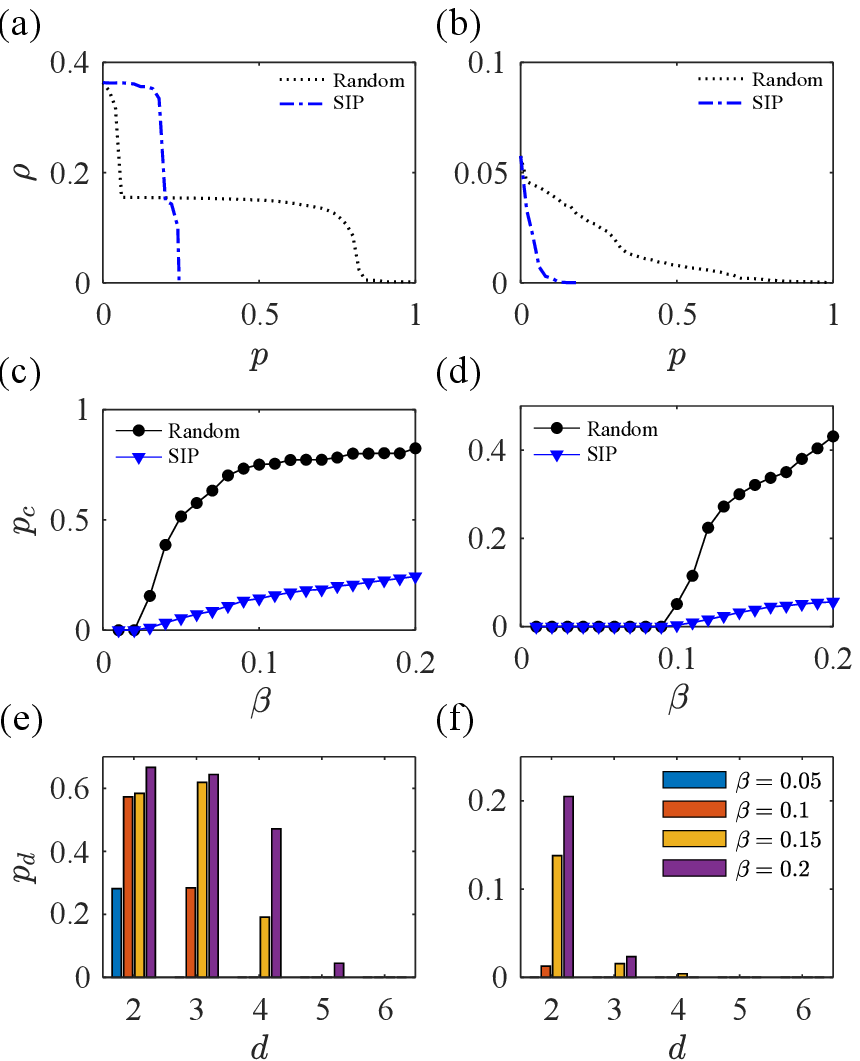}
		\caption{
         Random hyperedge immunization (Random) and SIP-based strategy tested in empirical hypergraphs: (a, c, e) the congressional bill cosponsorship (in 2000) hypergraph and (b, d, f) the protein interaction hypergraph.
        For nonuniform empirical hypergraphs, H-ES cannot be implemented due to the variety of hyperedge sizes, and EI is computationally inefficient due to the large hyperedges.
		(a, b) The density of infection $\rho$ versus the removed portion of the edges $p$.
        The recovery rate $\mu=0.2$ and the contagion rate $\beta = \beta_d = 0.2$ for all hyperedge sizes $d$.
        The efficient strategies generally exhibit a higher density of infection for small $p$, but herd immunity is achieved at lower HIT $p_c$, which is the minimally required portion of hyperedges that need to be immunized to eliminate the epidemics.
        (c, d) HIT $p_c$ as a function of the contagion rate $\beta = \beta_d$.
        Efficient epidemic containment is achieved by the SIP-based method with low computational cost.
        (e, f) The immunization rate of hyperedges with size $d$ plotted for various contagion rates.
        Small hyperedges are primarily targeted by the immunization strategy especially when $\beta$ is low.
		}
		\label{fig:plot_hypergraph}
	\end{figure}
    
   To test the immunization strategies, we implement the quasistationary method~\cite{Ferreira2011,Ferreira2012}, which is a standard simulation method used to study stationary states of stochastic processes with absorbing states.
   An absorbing state has zero probability of transitioning to other states.
   In this case, because both the contagion and recovery process involves an infected node, if all the nodes are in the susceptible state, it cannot turn into any other state: it is the absorbing state of the $s$-SIS model.
    The quasistationary method constrains the system in the active states.
    We keep track of a set of configurations of the system, which is referred to as the \textit{history}.
    With a certain probability, we replace one of the configurations in the history, randomly selected at each time step with the current state of the system.
    When the absorbing state is reached, the state of the system is replaced by a configuration randomly selected from the history.
    Here, we track 50 configurations and update with probability 0.2 at each time step.
    
    We first test the strategies in synthetic and empirical networks.
    These networks are selected as examples, and the relative effectiveness of the immunization strategies generally do not strongly vary from network to network.
    For the unclustered scale-free network, we use the static model~\cite{Goh2001c,Lee2006} with 5000 nodes, 15000 edges, and degree exponent 3.
    For the clustered scale-free network, we implement the model proposed in Ref.~\cite{Holme2002} with 5000 nodes, 15000 edges, degree exponent 3, and the parameter $p=0.8$, which makes the clustering coefficient 0.6.
    For empirical networks, we use the largest connected component of the airline network~\cite{Jia2019} which has 3354 nodes and 19162 edges.
    Each node represents an airport, and if there exists an airline between two airports, they are connected by an edge in the network.
    Another empirical network we use is the largest connected component of the general relativity and quantum cosmology collaboration network~\cite{Leskovec2007}.
    There are 4158 nodes and 13428 edges in the network.
    Each node represents an author of a paper submitted to the General Relativity and Cosmology category in arXiv, and if two authors coauthored a paper in the arXiv category from January 1993 to April 2003, they are connected by an edge in the network. 
    
    The results of the strategies in the networks are illustrated in Fig.~\ref{fig:plot_network}.
    We plot the density of infection versus the immunization rate $p$ for $\beta=\mu=0.2$ [Figs.~\ref{fig:plot_network}(a--d)].
    The density of infection of efficient strategies is often higher than that of random edge immunization for small immunization rate $p$, but for sufficiently large $p$, the density of infection drops quickly and achieves herd immunity at a lower $p_c$.
    The HIT $p_c$ is illustrated in Figs.~\ref{fig:plot_network}(e--h).
    One way to calculate the effective HIT of is to calculate the minimally required immunization rate to lower the density of infection below $1/N$.
    However, when simulating the stationary states of epidemic processes, if the system reaches its absorbing state, we arbitrarily adjust the system by reverting it back to one of its histories (quasistationary method) or activating a single site~\cite{Jo2020}. 
    Therefore, the state whose \textit{number} of infected nodes is close to zero is highly influenced by the choice of the simulation method which is not part of the epidemic model.
    To solve this problem, one can choose the herd immunity condition as the density of infection of 1\%, which is sometimes used as a threshold to be considered as \textit{subextensive} in networks~\cite{Im2018}.
    However, in real-world situations, an epidemic prevalence of 1\% is still an alerting scenario, and the epidemics cannot be considered under control.
    By choosing the density of infection of $\min \left(0.01, 1/\sqrt{N} \right)$ as the herd immunity condition, this dilemma can be resolved.
    In the thermodynamics limit $N\rightarrow \infty$, the epidemic density of the herd immunity condition converges to zero while the number of the infected nodes approaches infinity.

    The recovery rate is fixed to $\mu=0.2$.
    The HITs of the three efficient strategies are almost identical.
    To compare the HITs of efficient strategies more thoroughly, we plot the differences between the eigenscore strategy and two other strategies in Figs.~\ref{fig:plot_network}(i--l).
    The EI-based strategy is generally slightly more efficient than the eigenscore strategy, but it does not have an advantage over the SIP-based strategy, despite its higher computational cost. 
    Rather, the SIP-based strategy has a small advantage in the networks studied here, although the differences are marginal.
    
    Then, we test the strategies in 3-uniform hypergraphs.
    We use two synthetic models of 3-uniform hypergraphs: a static model with 2000 nodes, 4000 hyperedges, and degree exponent 3, and the $h$-PSO model introduced in Sec.~\ref{sec:hpso} with the same number of nodes, hyperedges, and degree exponent.
    The temperature $T=0.5$ and $R=1$ result in the clustering coefficient $C(H)=1.0430$.
    We illustrated the results in Fig.~\ref{fig:plot_uniform_hypergraph}.
    The density of infection $\rho$ of the strategies versus the immunization ratio $p$ for $\beta=\beta_3=\mu=0.2$ is depicted in Figs.~\ref{fig:plot_uniform_hypergraph}(a, b).
    The H-eigenscore and SIP-based method result in a higher density of infection for small immunization ratios, but eventually yield a smaller HIT $p_c$ for herd immunity [Figs.~\ref{fig:plot_uniform_hypergraph}(c, d)].
    The recovery rate is fixed to $\mu=0.2$.
    
    We test the SIP-based strategy in two empirical hypergraphs with various hyperedge sizes.
    One is the congressional bill cosponsorship hypergraph~\cite{Benson2018,Fowler2006}, which has 536 nodes and 2773 hyperedges whose mean size is 16.57 and maximum size is 323.
    Each node represents a US congressperson, and if a set of $d$ congresspeople cosponsored a bill in the year 2000, they are connected by a hyperedge of size $d$.
    The other is the protein interaction hypergraph~\cite{Klimm2020}, which has 8243 nodes and 6688 hyperedges whose mean size is 10.12 and maximum size is 421.
	Each node in the hypergraph represents a protein, and each hyperedge represents a type of multiprotein complex.
    Due to the large and heterogeneous size of hyperedges, only the SIP-based strategy can efficiently be implemented in these systems.
    We compare the density of infection of the strategy with random immunization in Figs.~\ref{fig:plot_hypergraph}(a, b).
    The recovery rate $\mu=0.2$ and the contagion rate for hyperedges are set $\beta_d=\beta=0.2$ independently of their sizes.
    While random immunization requires the majority of hyperedges to be immune to eliminate the epidemics, the SIP-based strategy achieves it with small $p_c$.
    The HITs are plotted for various contagion rates $\beta = \beta_d$ in Figs.~\ref{fig:plot_hypergraph}(c, d).
    The immunization rate of hyperedges of each size are illustrated in Fig. 3(e, f).
    Although removing large hyperedges affect large number of nodes, small hyperedges are primarily immunized especially when the contagion rate $\beta$ is low.
    This is because the nodes that are connected by a small hyperedge interact more strongly.
    It is interesting to point out that an epidemic containment strategy that immunizes groups in descending order of their size was effective in the localized regime~\cite{St-Onge2021} of higher-order epidemics~\cite{St-Onge2021a}.
  		
	\section{Conclusion\label{sec:conclusion}}
	
    In summary, we proposed an effective immunization strategy that immunizes hyperedges with high SIP that can be used in general hypergraphs, including networks.
    Hyperedges with high SIP are "hotspots" of the epidemics, and they can be identified and immunized.
	In case of information spreading processes, such hyperedges can be fostered to boost the information flow in the system. 
    We also show that H-eigenscore is a natural generalization of the eigenscore for hypergraphs.
    If all the hyperedges in a hypergraph have a size of 2, the H-eigenscore becomes identical to the eigenscore used in networks.
    Immunizing hyperedges with a high H-eigenscore effectively contains the epidemics, but the method can only be implemented in uniform hypergraphs.
    
    We tested the performance of the method and compared it with the state-of-the-art immunization strategy of the EI-based method in networks and hypergraphs. 
    In networks, the HIT $p_c$ of the SIP-based strategy is marginally smaller than that of the EI-based strategy, despite its lower computational cost.
    In hypergraphs, the SIP-based strategy yields significantly smaller HIT $p_c$ with lower computational cost.
	This suggests that SIP can serve as a centrality measure for hyperedges in general hypergraphs.
    The large disparity between the $p_c$ of an efficient immunization strategy and random immunization calls for scientific, data-driven, systematic policy-making for containment measures to eliminate epidemics with the minimum use of resources for vaccination and minimal regulation of air traffic and social gatherings.
    
    The IBMF used to calculate the SIP tend to overestimate the infection probability of the nodes (and, as a consequence, overestimate the global prevalence) because it ignores the correlations between the neighboring nodes.
    Recently introduced microscopic epidemic clique equations (MECLE)~\cite{Burgio2021}, which generalizes the epidemic-link equation to higher-order group interactions, predicted the density of infection and epidemic thresholds by taking the dynamic correlations between the neighboring nodes into account.
    An interesting work for the future might be to see how the performance of the SIP-based immunization strategy would be affected if the dynamical correlations are considered.
    Accounting for such correlations rapidly becomes unfeasible as the size of the hyperedges grow, therefore, it should be studied in hypergraphs whose hyperedges are not too large.

	\begin{acknowledgments}
		This research was supported by the NRF, Grant No.~NRF-2014R1A3A2069005. 
	\end{acknowledgments}
	

\begin{thebibliography}{84}%
\makeatletter
\providecommand \@ifxundefined [1]{%
 \@ifx{#1\undefined}
}%
\providecommand \@ifnum [1]{%
 \ifnum #1\expandafter \@firstoftwo
 \else \expandafter \@secondoftwo
 \fi
}%
\providecommand \@ifx [1]{%
 \ifx #1\expandafter \@firstoftwo
 \else \expandafter \@secondoftwo
 \fi
}%
\providecommand \natexlab [1]{#1}%
\providecommand \enquote  [1]{``#1''}%
\providecommand \bibnamefont  [1]{#1}%
\providecommand \bibfnamefont [1]{#1}%
\providecommand \citenamefont [1]{#1}%
\providecommand \href@noop [0]{\@secondoftwo}%
\providecommand \href [0]{\begingroup \@sanitize@url \@href}%
\providecommand \@href[1]{\@@startlink{#1}\@@href}%
\providecommand \@@href[1]{\endgroup#1\@@endlink}%
\providecommand \@sanitize@url [0]{\catcode `\\12\catcode `\$12\catcode
  `\&12\catcode `\#12\catcode `\^12\catcode `\_12\catcode `\%12\relax}%
\providecommand \@@startlink[1]{}%
\providecommand \@@endlink[0]{}%
\providecommand \url  [0]{\begingroup\@sanitize@url \@url }%
\providecommand \@url [1]{\endgroup\@href {#1}{\urlprefix }}%
\providecommand \urlprefix  [0]{URL }%
\providecommand \Eprint [0]{\href }%
\providecommand \doibase [0]{https://doi.org/}%
\providecommand \selectlanguage [0]{\@gobble}%
\providecommand \bibinfo  [0]{\@secondoftwo}%
\providecommand \bibfield  [0]{\@secondoftwo}%
\providecommand \translation [1]{[#1]}%
\providecommand \BibitemOpen [0]{}%
\providecommand \bibitemStop [0]{}%
\providecommand \bibitemNoStop [0]{.\EOS\space}%
\providecommand \EOS [0]{\spacefactor3000\relax}%
\providecommand \BibitemShut  [1]{\csname bibitem#1\endcsname}%
\let\auto@bib@innerbib\@empty
\bibitem [{\citenamefont {Dorogovtsev}\ \emph {et~al.}(2008)\citenamefont
  {Dorogovtsev}, \citenamefont {Goltsev},\ and\ \citenamefont
  {Mendes}}]{Dorogovtsev2008}%
  \BibitemOpen
  \bibfield  {author} {\bibinfo {author} {\bibfnamefont {S.~N.}\ \bibnamefont
  {Dorogovtsev}}, \bibinfo {author} {\bibfnamefont {A.~V.}\ \bibnamefont
  {Goltsev}},\ and\ \bibinfo {author} {\bibfnamefont {J.~F.~F.}\ \bibnamefont
  {Mendes}},\ }\bibfield  {title} {\bibinfo {title} {{Critical phenomena in
  complex networks}},\ }\href {https://doi.org/10.1103/RevModPhys.80.1275}
  {\bibfield  {journal} {\bibinfo  {journal} {Rev. Mod. Phys.}\ }\textbf
  {\bibinfo {volume} {80}},\ \bibinfo {pages} {1275} (\bibinfo {year}
  {2008})}\BibitemShut {NoStop}%
\bibitem [{\citenamefont {Zimmermann}\ \emph {et~al.}(2004)\citenamefont
  {Zimmermann}, \citenamefont {Egu\'{\i}luz},\ and\ \citenamefont
  {San~Miguel}}]{Zimmermann2004}%
  \BibitemOpen
  \bibfield  {author} {\bibinfo {author} {\bibfnamefont {M.~G.}\ \bibnamefont
  {Zimmermann}}, \bibinfo {author} {\bibfnamefont {V.~M.}\ \bibnamefont
  {Egu\'{\i}luz}},\ and\ \bibinfo {author} {\bibfnamefont {M.}~\bibnamefont
  {San~Miguel}},\ }\bibfield  {title} {\bibinfo {title} {Coevolution of
  dynamical states and interactions in dynamic networks},\ }\href
  {https://doi.org/10.1103/PhysRevE.69.065102} {\bibfield  {journal} {\bibinfo
  {journal} {Phys. Rev. E}\ }\textbf {\bibinfo {volume} {69}},\ \bibinfo
  {pages} {065102(R)} (\bibinfo {year} {2004})}\BibitemShut {NoStop}%
\bibitem [{\citenamefont {Dorogovtsev}\ and\ \citenamefont
  {Mendes}(2002)}]{Dorogovtsev2002}%
  \BibitemOpen
  \bibfield  {author} {\bibinfo {author} {\bibfnamefont {S.~N.}\ \bibnamefont
  {Dorogovtsev}}\ and\ \bibinfo {author} {\bibfnamefont {J.~F.}\ \bibnamefont
  {Mendes}},\ }\bibfield  {title} {\bibinfo {title} {{Evolution of networks}},\
  }\href {https://doi.org/10.1080/00018730110112519} {\bibfield  {journal}
  {\bibinfo  {journal} {Adv. Phys.}\ }\textbf {\bibinfo {volume} {51}},\
  \bibinfo {pages} {1079} (\bibinfo {year} {2002})}\BibitemShut {NoStop}%
\bibitem [{\citenamefont {Klemm}\ \emph {et~al.}(2003)\citenamefont {Klemm},
  \citenamefont {Egu\'{\i}luz}, \citenamefont {Toral},\ and\ \citenamefont
  {San~Miguel}}]{Klemm2003}%
  \BibitemOpen
  \bibfield  {author} {\bibinfo {author} {\bibfnamefont {K.}~\bibnamefont
  {Klemm}}, \bibinfo {author} {\bibfnamefont {V.~M.}\ \bibnamefont
  {Egu\'{\i}luz}}, \bibinfo {author} {\bibfnamefont {R.}~\bibnamefont
  {Toral}},\ and\ \bibinfo {author} {\bibfnamefont {M.}~\bibnamefont
  {San~Miguel}},\ }\bibfield  {title} {\bibinfo {title} {Nonequilibrium
  transitions in complex networks: A model of social interaction},\ }\href
  {https://doi.org/10.1103/PhysRevE.67.026120} {\bibfield  {journal} {\bibinfo
  {journal} {Phys. Rev. E}\ }\textbf {\bibinfo {volume} {67}},\ \bibinfo
  {pages} {026120} (\bibinfo {year} {2003})}\BibitemShut {NoStop}%
\bibitem [{\citenamefont {Nekovee}\ \emph {et~al.}(2007)\citenamefont
  {Nekovee}, \citenamefont {Moreno}, \citenamefont {Bianconi},\ and\
  \citenamefont {Marsili}}]{Nekovee2007}%
  \BibitemOpen
  \bibfield  {author} {\bibinfo {author} {\bibfnamefont {M.}~\bibnamefont
  {Nekovee}}, \bibinfo {author} {\bibfnamefont {Y.}~\bibnamefont {Moreno}},
  \bibinfo {author} {\bibfnamefont {G.}~\bibnamefont {Bianconi}},\ and\
  \bibinfo {author} {\bibfnamefont {M.}~\bibnamefont {Marsili}},\ }\bibfield
  {title} {\bibinfo {title} {{Theory of rumour spreading in complex social
  networks}},\ }\href {https://doi.org/10.1016/j.physa.2006.07.017} {\bibfield
  {journal} {\bibinfo  {journal} {Physica A}\ }\textbf {\bibinfo {volume}
  {374}},\ \bibinfo {pages} {457} (\bibinfo {year} {2007})}\BibitemShut
  {NoStop}%
\bibitem [{\citenamefont {Pastor-Satorras}\ \emph {et~al.}(2015)\citenamefont
  {Pastor-Satorras}, \citenamefont {Castellano}, \citenamefont {{Van
  Mieghem}},\ and\ \citenamefont {Vespignani}}]{Pastor-Satorras2015c}%
  \BibitemOpen
  \bibfield  {author} {\bibinfo {author} {\bibfnamefont {R.}~\bibnamefont
  {Pastor-Satorras}}, \bibinfo {author} {\bibfnamefont {C.}~\bibnamefont
  {Castellano}}, \bibinfo {author} {\bibfnamefont {P.}~\bibnamefont {{Van
  Mieghem}}},\ and\ \bibinfo {author} {\bibfnamefont {A.}~\bibnamefont
  {Vespignani}},\ }\bibfield  {title} {\bibinfo {title} {{Epidemic processes in
  complex networks}},\ }\href {https://doi.org/10.1103/RevModPhys.87.925}
  {\bibfield  {journal} {\bibinfo  {journal} {Rev. Mod. Phys.}\ }\textbf
  {\bibinfo {volume} {87}},\ \bibinfo {pages} {925} (\bibinfo {year}
  {2015})}\BibitemShut {NoStop}%
\bibitem [{\citenamefont {Katona}\ \emph {et~al.}(2011)\citenamefont {Katona},
  \citenamefont {Zubcsek},\ and\ \citenamefont {Sarvary}}]{Katona2011}%
  \BibitemOpen
  \bibfield  {author} {\bibinfo {author} {\bibfnamefont {Z.}~\bibnamefont
  {Katona}}, \bibinfo {author} {\bibfnamefont {P.~P.}\ \bibnamefont
  {Zubcsek}},\ and\ \bibinfo {author} {\bibfnamefont {M.}~\bibnamefont
  {Sarvary}},\ }\bibfield  {title} {\bibinfo {title} {{Network Effects and
  Personal Influences: The Diffusion of an Online Social Network}},\ }\href
  {https://doi.org/10.1509/jmkr.48.3.425} {\bibfield  {journal} {\bibinfo
  {journal} {J. Mark. Res.}\ }\textbf {\bibinfo {volume} {48}},\ \bibinfo
  {pages} {425} (\bibinfo {year} {2011})}\BibitemShut {NoStop}%
\bibitem [{\citenamefont {ROGERS}(2004)}]{Rogers2004}%
  \BibitemOpen
  \bibfield  {author} {\bibinfo {author} {\bibfnamefont {E.~M.}\ \bibnamefont
  {ROGERS}},\ }\bibfield  {title} {\bibinfo {title} {{A Prospective and
  Retrospective Look at the Diffusion Model}},\ }\href
  {https://doi.org/10.1080/10810730490271449} {\bibfield  {journal} {\bibinfo
  {journal} {J. Health Commun.}\ }\textbf {\bibinfo {volume} {9}},\ \bibinfo
  {pages} {13} (\bibinfo {year} {2004})}\BibitemShut {NoStop}%
\bibitem [{\citenamefont {Acemoğlu}\ \emph {et~al.}(2013)\citenamefont
  {Acemoğlu}, \citenamefont {Como}, \citenamefont {Fagnani},\ and\
  \citenamefont {Ozdaglar}}]{Acemoglu2013}%
  \BibitemOpen
  \bibfield  {author} {\bibinfo {author} {\bibfnamefont {D.}~\bibnamefont
  {Acemoğlu}}, \bibinfo {author} {\bibfnamefont {G.}~\bibnamefont {Como}},
  \bibinfo {author} {\bibfnamefont {F.}~\bibnamefont {Fagnani}},\ and\ \bibinfo
  {author} {\bibfnamefont {A.}~\bibnamefont {Ozdaglar}},\ }\bibfield  {title}
  {\bibinfo {title} {{Opinion Fluctuations and Disagreement in Social
  Networks}},\ }\href {https://doi.org/10.1287/moor.1120.0570} {\bibfield
  {journal} {\bibinfo  {journal} {Math. Oper. Res.}\ }\textbf {\bibinfo
  {volume} {38}},\ \bibinfo {pages} {1} (\bibinfo {year} {2013})}\BibitemShut
  {NoStop}%
\bibitem [{\citenamefont {Grabowski}\ and\ \citenamefont
  {Kosi{\'{n}}ski}(2006)}]{Grabowski2006}%
  \BibitemOpen
  \bibfield  {author} {\bibinfo {author} {\bibfnamefont {A.}~\bibnamefont
  {Grabowski}}\ and\ \bibinfo {author} {\bibfnamefont {R.}~\bibnamefont
  {Kosi{\'{n}}ski}},\ }\bibfield  {title} {\bibinfo {title} {{Ising-based model
  of opinion formation in a complex network of interpersonal interactions}},\
  }\href {https://doi.org/10.1016/j.physa.2005.06.102} {\bibfield  {journal}
  {\bibinfo  {journal} {Physica A}\ }\textbf {\bibinfo {volume} {361}},\
  \bibinfo {pages} {651} (\bibinfo {year} {2006})}\BibitemShut {NoStop}%
\bibitem [{\citenamefont {Watts}\ and\ \citenamefont
  {Dodds}(2007)}]{Watts2007}%
  \BibitemOpen
  \bibfield  {author} {\bibinfo {author} {\bibfnamefont {D.~J.}\ \bibnamefont
  {Watts}}\ and\ \bibinfo {author} {\bibfnamefont {P.~S.}\ \bibnamefont
  {Dodds}},\ }\bibfield  {title} {\bibinfo {title} {{Influentials, Networks,
  and Public Opinion Formation}},\ }\href {https://doi.org/10.1086/518527}
  {\bibfield  {journal} {\bibinfo  {journal} {J. Consum. Res.}\ }\textbf
  {\bibinfo {volume} {34}},\ \bibinfo {pages} {441} (\bibinfo {year}
  {2007})}\BibitemShut {NoStop}%
\bibitem [{\citenamefont {Chen}\ \emph {et~al.}(2020)\citenamefont {Chen},
  \citenamefont {Wang}, \citenamefont {Shen}, \citenamefont {Zhang},\ and\
  \citenamefont {Bianconi}}]{Chen2020}%
  \BibitemOpen
  \bibfield  {author} {\bibinfo {author} {\bibfnamefont {H.}~\bibnamefont
  {Chen}}, \bibinfo {author} {\bibfnamefont {S.}~\bibnamefont {Wang}}, \bibinfo
  {author} {\bibfnamefont {C.}~\bibnamefont {Shen}}, \bibinfo {author}
  {\bibfnamefont {H.}~\bibnamefont {Zhang}},\ and\ \bibinfo {author}
  {\bibfnamefont {G.}~\bibnamefont {Bianconi}},\ }\bibfield  {title} {\bibinfo
  {title} {{Non-Markovian majority-vote model}},\ }\href
  {https://doi.org/10.1103/PhysRevE.102.062311} {\bibfield  {journal} {\bibinfo
   {journal} {Phys. Rev. E}\ }\textbf {\bibinfo {volume} {102}},\ \bibinfo
  {pages} {062311} (\bibinfo {year} {2020})}\BibitemShut {NoStop}%
\bibitem [{\citenamefont {Hoang}\ and\ \citenamefont
  {Antoncic}(2003)}]{Hoang2003}%
  \BibitemOpen
  \bibfield  {author} {\bibinfo {author} {\bibfnamefont {H.}~\bibnamefont
  {Hoang}}\ and\ \bibinfo {author} {\bibfnamefont {B.}~\bibnamefont
  {Antoncic}},\ }\bibfield  {title} {\bibinfo {title} {{Network-based research
  in entrepreneurship}},\ }\href
  {https://doi.org/10.1016/S0883-9026(02)00081-2} {\bibfield  {journal}
  {\bibinfo  {journal} {J. Bus. Ventur.}\ }\textbf {\bibinfo {volume} {18}},\
  \bibinfo {pages} {165} (\bibinfo {year} {2003})}\BibitemShut {NoStop}%
\bibitem [{\citenamefont {Newman}(2003)}]{Newman2003}%
  \BibitemOpen
  \bibfield  {author} {\bibinfo {author} {\bibfnamefont {M.~E.~J.}\
  \bibnamefont {Newman}},\ }\bibfield  {title} {\bibinfo {title} {{The
  Structure and Function of Complex Networks}},\ }\href
  {https://doi.org/10.1137/S003614450342480} {\bibfield  {journal} {\bibinfo
  {journal} {SIAM Rev.}\ }\textbf {\bibinfo {volume} {45}},\ \bibinfo {pages}
  {167} (\bibinfo {year} {2003})}\BibitemShut {NoStop}%
\bibitem [{\citenamefont {Boccaletti}\ \emph {et~al.}(2006)\citenamefont
  {Boccaletti}, \citenamefont {Latora}, \citenamefont {Moreno}, \citenamefont
  {Chavez},\ and\ \citenamefont {Hwang}}]{Boccaletti2006}%
  \BibitemOpen
  \bibfield  {author} {\bibinfo {author} {\bibfnamefont {S.}~\bibnamefont
  {Boccaletti}}, \bibinfo {author} {\bibfnamefont {V.}~\bibnamefont {Latora}},
  \bibinfo {author} {\bibfnamefont {Y.}~\bibnamefont {Moreno}}, \bibinfo
  {author} {\bibfnamefont {M.}~\bibnamefont {Chavez}},\ and\ \bibinfo {author}
  {\bibfnamefont {D.}~\bibnamefont {Hwang}},\ }\bibfield  {title} {\bibinfo
  {title} {{Complex networks: Structure and dynamics}},\ }\href
  {https://doi.org/10.1016/j.physrep.2005.10.009} {\bibfield  {journal}
  {\bibinfo  {journal} {Phys. Rep.}\ }\textbf {\bibinfo {volume} {424}},\
  \bibinfo {pages} {175} (\bibinfo {year} {2006})}\BibitemShut {NoStop}%
\bibitem [{\citenamefont {Vel{\'{a}}squez-Rojas}\ \emph
  {et~al.}(2020)\citenamefont {Vel{\'{a}}squez-Rojas}, \citenamefont {Ventura},
  \citenamefont {Connaughton}, \citenamefont {Moreno}, \citenamefont
  {Rodrigues},\ and\ \citenamefont {Vazquez}}]{Velasquez-Rojas2020}%
  \BibitemOpen
  \bibfield  {author} {\bibinfo {author} {\bibfnamefont {F.}~\bibnamefont
  {Vel{\'{a}}squez-Rojas}}, \bibinfo {author} {\bibfnamefont {P.~C.}\
  \bibnamefont {Ventura}}, \bibinfo {author} {\bibfnamefont {C.}~\bibnamefont
  {Connaughton}}, \bibinfo {author} {\bibfnamefont {Y.}~\bibnamefont {Moreno}},
  \bibinfo {author} {\bibfnamefont {F.~A.}\ \bibnamefont {Rodrigues}},\ and\
  \bibinfo {author} {\bibfnamefont {F.}~\bibnamefont {Vazquez}},\ }\bibfield
  {title} {\bibinfo {title} {{Disease and information spreading at different
  speeds in multiplex networks}},\ }\href
  {https://doi.org/10.1103/PhysRevE.102.022312} {\bibfield  {journal} {\bibinfo
   {journal} {Phys. Rev. E}\ }\textbf {\bibinfo {volume} {102}},\ \bibinfo
  {pages} {022312} (\bibinfo {year} {2020})}\BibitemShut {NoStop}%
\bibitem [{\citenamefont {Jia}\ and\ \citenamefont {Benson}(2019)}]{Jia2019}%
  \BibitemOpen
  \bibfield  {author} {\bibinfo {author} {\bibfnamefont {J.}~\bibnamefont
  {Jia}}\ and\ \bibinfo {author} {\bibfnamefont {A.~R.}\ \bibnamefont
  {Benson}},\ }\bibfield  {title} {\bibinfo {title} {{Random Spatial Network
  Models for Core-Periphery Structure}},\ }in\ \href
  {https://doi.org/10.1145/3289600.3290976} {\emph {\bibinfo {booktitle} {Proc.
  Twelfth ACM Int. Conf. Web Search Data Min.}}}\ (\bibinfo  {publisher}
  {ACM},\ \bibinfo {address} {New York, NY, USA},\ \bibinfo {year} {2019})\
  pp.\ \bibinfo {pages} {366--374}\BibitemShut {NoStop}%
\bibitem [{\citenamefont {Leskovec}\ \emph {et~al.}(2007)\citenamefont
  {Leskovec}, \citenamefont {Kleinberg},\ and\ \citenamefont
  {Faloutsos}}]{Leskovec2007}%
  \BibitemOpen
  \bibfield  {author} {\bibinfo {author} {\bibfnamefont {J.}~\bibnamefont
  {Leskovec}}, \bibinfo {author} {\bibfnamefont {J.}~\bibnamefont
  {Kleinberg}},\ and\ \bibinfo {author} {\bibfnamefont {C.}~\bibnamefont
  {Faloutsos}},\ }\bibfield  {title} {\bibinfo {title} {{Graph evolution}},\
  }\href {https://doi.org/10.1145/1217299.1217301} {\bibfield  {journal}
  {\bibinfo  {journal} {ACM Trans. Knowl. Discov. Data}\ }\textbf {\bibinfo
  {volume} {1}},\ \bibinfo {pages} {2} (\bibinfo {year} {2007})}\BibitemShut
  {NoStop}%
\bibitem [{\citenamefont {Fowler}(2006)}]{Fowler2006}%
  \BibitemOpen
  \bibfield  {author} {\bibinfo {author} {\bibfnamefont {J.~H.}\ \bibnamefont
  {Fowler}},\ }\bibfield  {title} {\bibinfo {title} {{Connecting the Congress:
  A Study of Cosponsorship Networks}},\ }\href
  {https://doi.org/10.1093/pan/mpl002} {\bibfield  {journal} {\bibinfo
  {journal} {Polit. Anal.}\ }\textbf {\bibinfo {volume} {14}},\ \bibinfo
  {pages} {456} (\bibinfo {year} {2006})}\BibitemShut {NoStop}%
\bibitem [{\citenamefont {Pastor-Satorras}\ and\ \citenamefont
  {Vespignani}(2001{\natexlab{a}})}]{Pastor-Satorras2001}%
  \BibitemOpen
  \bibfield  {author} {\bibinfo {author} {\bibfnamefont {R.}~\bibnamefont
  {Pastor-Satorras}}\ and\ \bibinfo {author} {\bibfnamefont {A.}~\bibnamefont
  {Vespignani}},\ }\bibfield  {title} {\bibinfo {title} {{Epidemic spreading in
  scale-free networks}},\ }\href {https://doi.org/10.1103/PhysRevLett.86.3200}
  {\bibfield  {journal} {\bibinfo  {journal} {Phys. Rev. Lett.}\ }\textbf
  {\bibinfo {volume} {86}},\ \bibinfo {pages} {3200} (\bibinfo {year}
  {2001}{\natexlab{a}})}\BibitemShut {NoStop}%
\bibitem [{\citenamefont {Moreno}\ \emph {et~al.}(2002)\citenamefont {Moreno},
  \citenamefont {Pastor-Satorras},\ and\ \citenamefont
  {Vespignani}}]{Moreno2002a}%
  \BibitemOpen
  \bibfield  {author} {\bibinfo {author} {\bibfnamefont {Y.}~\bibnamefont
  {Moreno}}, \bibinfo {author} {\bibfnamefont {R.}~\bibnamefont
  {Pastor-Satorras}},\ and\ \bibinfo {author} {\bibfnamefont {A.}~\bibnamefont
  {Vespignani}},\ }\bibfield  {title} {\bibinfo {title} {{Epidemic outbreaks in
  complex heterogeneous networks}},\ }\href
  {https://doi.org/10.1140/epjb/e20020122} {\bibfield  {journal} {\bibinfo
  {journal} {Eur. Phys. J. B}\ }\textbf {\bibinfo {volume} {26}},\ \bibinfo
  {pages} {521} (\bibinfo {year} {2002})}\BibitemShut {NoStop}%
\bibitem [{\citenamefont {Bar}\ and\ \citenamefont {Mukamel}(2014)}]{Bar2014}%
  \BibitemOpen
  \bibfield  {author} {\bibinfo {author} {\bibfnamefont {A.}~\bibnamefont
  {Bar}}\ and\ \bibinfo {author} {\bibfnamefont {D.}~\bibnamefont {Mukamel}},\
  }\bibfield  {title} {\bibinfo {title} {{Mixed-Order Phase Transition in a
  One-Dimensional Model}},\ }\href
  {https://doi.org/10.1103/PhysRevLett.112.015701} {\bibfield  {journal}
  {\bibinfo  {journal} {Phys. Rev. Lett.}\ }\textbf {\bibinfo {volume} {112}},\
  \bibinfo {pages} {015701} (\bibinfo {year} {2014})}\BibitemShut {NoStop}%
\bibitem [{\citenamefont {Choi}\ \emph {et~al.}(2017)\citenamefont {Choi},
  \citenamefont {Lee},\ and\ \citenamefont {Kahng}}]{Choi2017}%
  \BibitemOpen
  \bibfield  {author} {\bibinfo {author} {\bibfnamefont {W.}~\bibnamefont
  {Choi}}, \bibinfo {author} {\bibfnamefont {D.}~\bibnamefont {Lee}},\ and\
  \bibinfo {author} {\bibfnamefont {B.}~\bibnamefont {Kahng}},\ }\bibfield
  {title} {\bibinfo {title} {{Mixed-order phase transition in a two-step
  contagion model with a single infectious seed}},\ }\href
  {https://doi.org/10.1103/PhysRevE.95.022304} {\bibfield  {journal} {\bibinfo
  {journal} {Phys. Rev. E}\ }\textbf {\bibinfo {volume} {95}},\ \bibinfo
  {pages} {022304} (\bibinfo {year} {2017})}\BibitemShut {NoStop}%
\bibitem [{\citenamefont {Lee}\ \emph {et~al.}(2017)\citenamefont {Lee},
  \citenamefont {Choi}, \citenamefont {Kert{\'{e}}sz},\ and\ \citenamefont
  {Kahng}}]{Lee2017}%
  \BibitemOpen
  \bibfield  {author} {\bibinfo {author} {\bibfnamefont {D.}~\bibnamefont
  {Lee}}, \bibinfo {author} {\bibfnamefont {W.}~\bibnamefont {Choi}}, \bibinfo
  {author} {\bibfnamefont {J.}~\bibnamefont {Kert{\'{e}}sz}},\ and\ \bibinfo
  {author} {\bibfnamefont {B.}~\bibnamefont {Kahng}},\ }\bibfield  {title}
  {\bibinfo {title} {{Universal mechanism for hybrid percolation
  transitions}},\ }\href {https://doi.org/10.1038/s41598-017-06182-3}
  {\bibfield  {journal} {\bibinfo  {journal} {Sci. Rep.}\ }\textbf {\bibinfo
  {volume} {7}},\ \bibinfo {pages} {5723} (\bibinfo {year} {2017})}\BibitemShut
  {NoStop}%
\bibitem [{\citenamefont {Lambiotte}\ \emph {et~al.}(2019)\citenamefont
  {Lambiotte}, \citenamefont {Rosvall},\ and\ \citenamefont
  {Scholtes}}]{Lambiotte2019}%
  \BibitemOpen
  \bibfield  {author} {\bibinfo {author} {\bibfnamefont {R.}~\bibnamefont
  {Lambiotte}}, \bibinfo {author} {\bibfnamefont {M.}~\bibnamefont {Rosvall}},\
  and\ \bibinfo {author} {\bibfnamefont {I.}~\bibnamefont {Scholtes}},\
  }\bibfield  {title} {\bibinfo {title} {{From networks to optimal higher-order
  models of complex systems}},\ }\href
  {https://doi.org/10.1038/s41567-019-0459-y} {\bibfield  {journal} {\bibinfo
  {journal} {Nat. Phys.}\ }\textbf {\bibinfo {volume} {15}},\ \bibinfo {pages}
  {313} (\bibinfo {year} {2019})}\BibitemShut {NoStop}%
\bibitem [{\citenamefont {Battiston}\ \emph {et~al.}(2020)\citenamefont
  {Battiston}, \citenamefont {Cencetti}, \citenamefont {Iacopini},
  \citenamefont {Latora}, \citenamefont {Lucas}, \citenamefont {Patania},
  \citenamefont {Young},\ and\ \citenamefont {Petri}}]{Battiston2020}%
  \BibitemOpen
  \bibfield  {author} {\bibinfo {author} {\bibfnamefont {F.}~\bibnamefont
  {Battiston}}, \bibinfo {author} {\bibfnamefont {G.}~\bibnamefont {Cencetti}},
  \bibinfo {author} {\bibfnamefont {I.}~\bibnamefont {Iacopini}}, \bibinfo
  {author} {\bibfnamefont {V.}~\bibnamefont {Latora}}, \bibinfo {author}
  {\bibfnamefont {M.}~\bibnamefont {Lucas}}, \bibinfo {author} {\bibfnamefont
  {A.}~\bibnamefont {Patania}}, \bibinfo {author} {\bibfnamefont {J.-G.}\
  \bibnamefont {Young}},\ and\ \bibinfo {author} {\bibfnamefont
  {G.}~\bibnamefont {Petri}},\ }\bibfield  {title} {\bibinfo {title} {{Networks
  beyond pairwise interactions: Structure and dynamics}},\ }\href
  {https://doi.org/10.1016/j.physrep.2020.05.004} {\bibfield  {journal}
  {\bibinfo  {journal} {Phys. Rep.}\ }\textbf {\bibinfo {volume} {874}},\
  \bibinfo {pages} {1} (\bibinfo {year} {2020})}\BibitemShut {NoStop}%
\bibitem [{\citenamefont {Bianconi}\ \emph {et~al.}(2021)\citenamefont
  {Bianconi}, \citenamefont {Sun}, \citenamefont {Rapisardi},\ and\
  \citenamefont {Arenas}}]{Bianconi2021}%
  \BibitemOpen
  \bibfield  {author} {\bibinfo {author} {\bibfnamefont {G.}~\bibnamefont
  {Bianconi}}, \bibinfo {author} {\bibfnamefont {H.}~\bibnamefont {Sun}},
  \bibinfo {author} {\bibfnamefont {G.}~\bibnamefont {Rapisardi}},\ and\
  \bibinfo {author} {\bibfnamefont {A.}~\bibnamefont {Arenas}},\ }\bibfield
  {title} {\bibinfo {title} {{Message-passing approach to epidemic tracing and
  mitigation with apps}},\ }\href
  {https://doi.org/10.1103/PhysRevResearch.3.L012014} {\bibfield  {journal}
  {\bibinfo  {journal} {Phys. Rev. Research}\ }\textbf {\bibinfo {volume}
  {3}},\ \bibinfo {pages} {L012014} (\bibinfo {year} {2021})}\BibitemShut
  {NoStop}%
\bibitem [{\citenamefont {{Hern{\'{a}}ndez Serrano}}\ and\ \citenamefont
  {{S{\'{a}}nchez G{\'{o}}mez}}(2020)}]{HernandezSerrano2020}%
  \BibitemOpen
  \bibfield  {author} {\bibinfo {author} {\bibfnamefont {D.}~\bibnamefont
  {{Hern{\'{a}}ndez Serrano}}}\ and\ \bibinfo {author} {\bibfnamefont
  {D.}~\bibnamefont {{S{\'{a}}nchez G{\'{o}}mez}}},\ }\bibfield  {title}
  {\bibinfo {title} {{Centrality measures in simplicial complexes: Applications
  of topological data analysis to network science}},\ }\href
  {https://doi.org/10.1016/j.amc.2020.125331} {\bibfield  {journal} {\bibinfo
  {journal} {Appl. Math. Comput.}\ }\textbf {\bibinfo {volume} {382}},\
  \bibinfo {pages} {125331} (\bibinfo {year} {2020})}\BibitemShut {NoStop}%
\bibitem [{\citenamefont {Patania}\ \emph {et~al.}(2017)\citenamefont
  {Patania}, \citenamefont {Petri},\ and\ \citenamefont
  {Vaccarino}}]{Patania2017}%
  \BibitemOpen
  \bibfield  {author} {\bibinfo {author} {\bibfnamefont {A.}~\bibnamefont
  {Patania}}, \bibinfo {author} {\bibfnamefont {G.}~\bibnamefont {Petri}},\
  and\ \bibinfo {author} {\bibfnamefont {F.}~\bibnamefont {Vaccarino}},\
  }\bibfield  {title} {\bibinfo {title} {{The shape of collaborations}},\
  }\href {https://doi.org/10.1140/epjds/s13688-017-0114-8} {\bibfield
  {journal} {\bibinfo  {journal} {EPJ Data Sci.}\ }\textbf {\bibinfo {volume}
  {6}},\ \bibinfo {pages} {18} (\bibinfo {year} {2017})}\BibitemShut {NoStop}%
\bibitem [{\citenamefont {Benson}\ \emph {et~al.}(2018)\citenamefont {Benson},
  \citenamefont {Abebe}, \citenamefont {Schaub}, \citenamefont {Jadbabaie},\
  and\ \citenamefont {Kleinberg}}]{Benson2018}%
  \BibitemOpen
  \bibfield  {author} {\bibinfo {author} {\bibfnamefont {A.~R.}\ \bibnamefont
  {Benson}}, \bibinfo {author} {\bibfnamefont {R.}~\bibnamefont {Abebe}},
  \bibinfo {author} {\bibfnamefont {M.~T.}\ \bibnamefont {Schaub}}, \bibinfo
  {author} {\bibfnamefont {A.}~\bibnamefont {Jadbabaie}},\ and\ \bibinfo
  {author} {\bibfnamefont {J.}~\bibnamefont {Kleinberg}},\ }\bibfield  {title}
  {\bibinfo {title} {{Simplicial closure and higher-order link prediction}},\
  }\href {https://doi.org/10.1073/pnas.1800683115} {\bibfield  {journal}
  {\bibinfo  {journal} {Proc. Natl. Acad. Sci. U.S.A.}\ }\textbf {\bibinfo
  {volume} {115}},\ \bibinfo {pages} {E11221} (\bibinfo {year}
  {2018})}\BibitemShut {NoStop}%
\bibitem [{\citenamefont {Petri}\ \emph {et~al.}(2014)\citenamefont {Petri},
  \citenamefont {Expert}, \citenamefont {Turkheimer}, \citenamefont
  {Carhart-Harris}, \citenamefont {Nutt}, \citenamefont {Hellyer},\ and\
  \citenamefont {Vaccarino}}]{Petri2014}%
  \BibitemOpen
  \bibfield  {author} {\bibinfo {author} {\bibfnamefont {G.}~\bibnamefont
  {Petri}}, \bibinfo {author} {\bibfnamefont {P.}~\bibnamefont {Expert}},
  \bibinfo {author} {\bibfnamefont {F.}~\bibnamefont {Turkheimer}}, \bibinfo
  {author} {\bibfnamefont {R.}~\bibnamefont {Carhart-Harris}}, \bibinfo
  {author} {\bibfnamefont {D.}~\bibnamefont {Nutt}}, \bibinfo {author}
  {\bibfnamefont {P.~J.}\ \bibnamefont {Hellyer}},\ and\ \bibinfo {author}
  {\bibfnamefont {F.}~\bibnamefont {Vaccarino}},\ }\bibfield  {title} {\bibinfo
  {title} {{Homological scaffolds of brain functional networks}},\ }\href
  {https://doi.org/10.1098/rsif.2014.0873} {\bibfield  {journal} {\bibinfo
  {journal} {J. R. Soc. Interface}\ }\textbf {\bibinfo {volume} {11}},\
  \bibinfo {pages} {20140873} (\bibinfo {year} {2014})}\BibitemShut {NoStop}%
\bibitem [{\citenamefont {Klimm}\ \emph {et~al.}(2021)\citenamefont {Klimm},
  \citenamefont {Deane},\ and\ \citenamefont {Reinert}}]{Klimm2020}%
  \BibitemOpen
  \bibfield  {author} {\bibinfo {author} {\bibfnamefont {F.}~\bibnamefont
  {Klimm}}, \bibinfo {author} {\bibfnamefont {C.~M.}\ \bibnamefont {Deane}},\
  and\ \bibinfo {author} {\bibfnamefont {G.}~\bibnamefont {Reinert}},\ }\href
  {https://doi.org/10.1093/comnet/cnaa028} {\bibfield  {journal} {\bibinfo
  {journal} {J. Complex Networks}\ }\textbf {\bibinfo {volume} {9}},\ \bibinfo
  {pages} {2020.04.03.023937} (\bibinfo {year} {2021})}\BibitemShut {NoStop}%
\bibitem [{\citenamefont {Burgio}\ \emph {et~al.}(2020)\citenamefont {Burgio},
  \citenamefont {Matamalas}, \citenamefont {G{\'{o}}mez},\ and\ \citenamefont
  {Arenas}}]{Burgio2020}%
  \BibitemOpen
  \bibfield  {author} {\bibinfo {author} {\bibfnamefont {G.}~\bibnamefont
  {Burgio}}, \bibinfo {author} {\bibfnamefont {J.~T.}\ \bibnamefont
  {Matamalas}}, \bibinfo {author} {\bibfnamefont {S.}~\bibnamefont
  {G{\'{o}}mez}},\ and\ \bibinfo {author} {\bibfnamefont {A.}~\bibnamefont
  {Arenas}},\ }\bibfield  {title} {\bibinfo {title} {{Evolution of Cooperation
  in the Presence of Higher-Order Interactions: From Networks to
  Hypergraphs}},\ }\href {https://doi.org/10.3390/e22070744} {\bibfield
  {journal} {\bibinfo  {journal} {Entropy}\ }\textbf {\bibinfo {volume} {22}},\
  \bibinfo {pages} {744} (\bibinfo {year} {2020})}\BibitemShut {NoStop}%
\bibitem [{\citenamefont {Alvarez-Rodriguez}\ \emph {et~al.}(2021)\citenamefont
  {Alvarez-Rodriguez}, \citenamefont {Battiston}, \citenamefont {de~Arruda},
  \citenamefont {Moreno}, \citenamefont {Perc},\ and\ \citenamefont
  {Latora}}]{Alvarez-Rodriguez2021}%
  \BibitemOpen
  \bibfield  {author} {\bibinfo {author} {\bibfnamefont {U.}~\bibnamefont
  {Alvarez-Rodriguez}}, \bibinfo {author} {\bibfnamefont {F.}~\bibnamefont
  {Battiston}}, \bibinfo {author} {\bibfnamefont {G.~F.}\ \bibnamefont
  {de~Arruda}}, \bibinfo {author} {\bibfnamefont {Y.}~\bibnamefont {Moreno}},
  \bibinfo {author} {\bibfnamefont {M.}~\bibnamefont {Perc}},\ and\ \bibinfo
  {author} {\bibfnamefont {V.}~\bibnamefont {Latora}},\ }\bibfield  {title}
  {\bibinfo {title} {{Evolutionary dynamics of higher-order interactions in
  social networks}},\ }\bibfield  {journal} {\bibinfo  {journal} {Nat. Hum.
  Behav.}\ }\href {https://doi.org/10.1038/s41562-020-01024-1}
  {10.1038/s41562-020-01024-1} (\bibinfo {year} {2021})\BibitemShut {NoStop}%
\bibitem [{\citenamefont {Carletti}\ \emph {et~al.}(2020)\citenamefont
  {Carletti}, \citenamefont {Battiston}, \citenamefont {Cencetti},\ and\
  \citenamefont {Fanelli}}]{Carletti2020}%
  \BibitemOpen
  \bibfield  {author} {\bibinfo {author} {\bibfnamefont {T.}~\bibnamefont
  {Carletti}}, \bibinfo {author} {\bibfnamefont {F.}~\bibnamefont {Battiston}},
  \bibinfo {author} {\bibfnamefont {G.}~\bibnamefont {Cencetti}},\ and\
  \bibinfo {author} {\bibfnamefont {D.}~\bibnamefont {Fanelli}},\ }\bibfield
  {title} {\bibinfo {title} {{Random walks on hypergraphs}},\ }\href
  {https://doi.org/10.1103/PhysRevE.101.022308} {\bibfield  {journal} {\bibinfo
   {journal} {Phys. Rev. E}\ }\textbf {\bibinfo {volume} {101}},\ \bibinfo
  {pages} {022308} (\bibinfo {year} {2020})}\BibitemShut {NoStop}%
\bibitem [{\citenamefont {St-Onge}\ \emph
  {et~al.}(2021{\natexlab{a}})\citenamefont {St-Onge}, \citenamefont
  {Thibeault}, \citenamefont {Allard}, \citenamefont {Dub{\'{e}}},\ and\
  \citenamefont {H{\'{e}}bert-Dufresne}}]{St-Onge2021}%
  \BibitemOpen
  \bibfield  {author} {\bibinfo {author} {\bibfnamefont {G.}~\bibnamefont
  {St-Onge}}, \bibinfo {author} {\bibfnamefont {V.}~\bibnamefont {Thibeault}},
  \bibinfo {author} {\bibfnamefont {A.}~\bibnamefont {Allard}}, \bibinfo
  {author} {\bibfnamefont {L.~J.}\ \bibnamefont {Dub{\'{e}}}},\ and\ \bibinfo
  {author} {\bibfnamefont {L.}~\bibnamefont {H{\'{e}}bert-Dufresne}},\
  }\bibfield  {title} {\bibinfo {title} {{Master equation analysis of
  mesoscopic localization in contagion dynamics on higher-order networks}},\
  }\href {https://doi.org/10.1103/PhysRevE.103.032301} {\bibfield  {journal}
  {\bibinfo  {journal} {Phys. Rev. E}\ }\textbf {\bibinfo {volume} {103}},\
  \bibinfo {pages} {032301} (\bibinfo {year} {2021}{\natexlab{a}})}\BibitemShut
  {NoStop}%
\bibitem [{\citenamefont {{Ferraz de Arruda}}\ \emph
  {et~al.}(2021)\citenamefont {{Ferraz de Arruda}}, \citenamefont {Tizzani},\
  and\ \citenamefont {Moreno}}]{DeArruda2020}%
  \BibitemOpen
  \bibfield  {author} {\bibinfo {author} {\bibfnamefont {G.}~\bibnamefont
  {{Ferraz de Arruda}}}, \bibinfo {author} {\bibfnamefont {M.}~\bibnamefont
  {Tizzani}},\ and\ \bibinfo {author} {\bibfnamefont {Y.}~\bibnamefont
  {Moreno}},\ }\bibfield  {title} {\bibinfo {title} {{Phase transitions and
  stability of dynamical processes on hypergraphs}},\ }\href
  {https://doi.org/10.1038/s42005-021-00525-3} {\bibfield  {journal} {\bibinfo
  {journal} {Commun. Phys.}\ }\textbf {\bibinfo {volume} {4}},\ \bibinfo
  {pages} {24} (\bibinfo {year} {2021})}\BibitemShut {NoStop}%
\bibitem [{\citenamefont {Lee}\ \emph {et~al.}(2021)\citenamefont {Lee},
  \citenamefont {Lee}, \citenamefont {Oh}, \citenamefont {Lee},\ and\
  \citenamefont {Kahng}}]{Lee2021}%
  \BibitemOpen
  \bibfield  {author} {\bibinfo {author} {\bibfnamefont {Y.}~\bibnamefont
  {Lee}}, \bibinfo {author} {\bibfnamefont {J.}~\bibnamefont {Lee}}, \bibinfo
  {author} {\bibfnamefont {S.~M.}\ \bibnamefont {Oh}}, \bibinfo {author}
  {\bibfnamefont {D.}~\bibnamefont {Lee}},\ and\ \bibinfo {author}
  {\bibfnamefont {B.}~\bibnamefont {Kahng}},\ }\bibfield  {title} {\bibinfo
  {title} {{Homological percolation transitions in growing simplicial
  complexes}},\ }\href {https://doi.org/10.1063/5.0047608} {\bibfield
  {journal} {\bibinfo  {journal} {Chaos}\ }\textbf {\bibinfo {volume} {31}},\
  \bibinfo {pages} {041102} (\bibinfo {year} {2021})}\BibitemShut {NoStop}%
\bibitem [{\citenamefont {Iacopini}\ \emph {et~al.}(2019)\citenamefont
  {Iacopini}, \citenamefont {Petri}, \citenamefont {Barrat},\ and\
  \citenamefont {Latora}}]{Iacopini2019}%
  \BibitemOpen
  \bibfield  {author} {\bibinfo {author} {\bibfnamefont {I.}~\bibnamefont
  {Iacopini}}, \bibinfo {author} {\bibfnamefont {G.}~\bibnamefont {Petri}},
  \bibinfo {author} {\bibfnamefont {A.}~\bibnamefont {Barrat}},\ and\ \bibinfo
  {author} {\bibfnamefont {V.}~\bibnamefont {Latora}},\ }\bibfield  {title}
  {\bibinfo {title} {{Simplicial models of social contagion}},\ }\href
  {https://doi.org/10.1038/s41467-019-10431-6} {\bibfield  {journal} {\bibinfo
  {journal} {Nat. Commun.}\ }\textbf {\bibinfo {volume} {10}},\ \bibinfo
  {pages} {2485} (\bibinfo {year} {2019})}\BibitemShut {NoStop}%
\bibitem [{\citenamefont {Jhun}\ \emph {et~al.}(2019)\citenamefont {Jhun},
  \citenamefont {Jo},\ and\ \citenamefont {Kahng}}]{Jhun2019a}%
  \BibitemOpen
  \bibfield  {author} {\bibinfo {author} {\bibfnamefont {B.}~\bibnamefont
  {Jhun}}, \bibinfo {author} {\bibfnamefont {M.}~\bibnamefont {Jo}},\ and\
  \bibinfo {author} {\bibfnamefont {B.}~\bibnamefont {Kahng}},\ }\bibfield
  {title} {\bibinfo {title} {{Simplicial SIS model in scale-free uniform
  hypergraph}},\ }\href {https://doi.org/10.1088/1742-5468/ab5367} {\bibfield
  {journal} {\bibinfo  {journal} {J. Stat. Mech.}\ }\textbf {\bibinfo {volume}
  {2019}},\ \bibinfo {pages} {123207} (\bibinfo {year} {2019})}\BibitemShut
  {NoStop}%
\bibitem [{\citenamefont {Matamalas}\ \emph {et~al.}(2020)\citenamefont
  {Matamalas}, \citenamefont {G\'omez},\ and\ \citenamefont
  {Arenas}}]{Matamalas2020}%
  \BibitemOpen
  \bibfield  {author} {\bibinfo {author} {\bibfnamefont {J.~T.}\ \bibnamefont
  {Matamalas}}, \bibinfo {author} {\bibfnamefont {S.}~\bibnamefont {G\'omez}},\
  and\ \bibinfo {author} {\bibfnamefont {A.}~\bibnamefont {Arenas}},\
  }\bibfield  {title} {\bibinfo {title} {Abrupt phase transition of epidemic
  spreading in simplicial complexes},\ }\href
  {https://doi.org/10.1103/PhysRevResearch.2.012049} {\bibfield  {journal}
  {\bibinfo  {journal} {Phys. Rev. Research}\ }\textbf {\bibinfo {volume}
  {2}},\ \bibinfo {pages} {012049(R)} (\bibinfo {year} {2020})}\BibitemShut
  {NoStop}%
\bibitem [{\citenamefont {Landry}\ and\ \citenamefont
  {Restrepo}(2020)}]{Landry2020}%
  \BibitemOpen
  \bibfield  {author} {\bibinfo {author} {\bibfnamefont {N.~W.}\ \bibnamefont
  {Landry}}\ and\ \bibinfo {author} {\bibfnamefont {J.~G.}\ \bibnamefont
  {Restrepo}},\ }\bibfield  {title} {\bibinfo {title} {{The effect of
  heterogeneity on hypergraph contagion models}},\ }\href
  {https://doi.org/10.1063/5.0020034} {\bibfield  {journal} {\bibinfo
  {journal} {Chaos}\ }\textbf {\bibinfo {volume} {30}},\ \bibinfo {pages}
  {103117} (\bibinfo {year} {2020})}\BibitemShut {NoStop}%
\bibitem [{\citenamefont {Chowdhary}\ \emph {et~al.}(2021)\citenamefont
  {Chowdhary}, \citenamefont {Kumar}, \citenamefont {Cencetti}, \citenamefont
  {Iacopini},\ and\ \citenamefont {Battiston}}]{Chowdhary2021}%
  \BibitemOpen
  \bibfield  {author} {\bibinfo {author} {\bibfnamefont {S.}~\bibnamefont
  {Chowdhary}}, \bibinfo {author} {\bibfnamefont {A.}~\bibnamefont {Kumar}},
  \bibinfo {author} {\bibfnamefont {G.}~\bibnamefont {Cencetti}}, \bibinfo
  {author} {\bibfnamefont {I.}~\bibnamefont {Iacopini}},\ and\ \bibinfo
  {author} {\bibfnamefont {F.}~\bibnamefont {Battiston}},\ }\href
  {https://doi.org/10.1088/2632-072X/ac12bd} {\bibfield  {journal} {\bibinfo
  {journal} {J. Phys. Complex.}\ }\textbf {\bibinfo {volume} {2}},\ \bibinfo
  {pages} {035019} (\bibinfo {year} {2021})},\ \Eprint
  {https://arxiv.org/abs/2105.04455} {arXiv:2105.04455} \BibitemShut {NoStop}%
\bibitem [{\citenamefont {Wang}\ \emph {et~al.}(2021)\citenamefont {Wang},
  \citenamefont {Zhao}, \citenamefont {Luo},\ and\ \citenamefont
  {Leng}}]{Wang2021}%
  \BibitemOpen
  \bibfield  {author} {\bibinfo {author} {\bibfnamefont {D.}~\bibnamefont
  {Wang}}, \bibinfo {author} {\bibfnamefont {Y.}~\bibnamefont {Zhao}}, \bibinfo
  {author} {\bibfnamefont {J.}~\bibnamefont {Luo}},\ and\ \bibinfo {author}
  {\bibfnamefont {H.}~\bibnamefont {Leng}},\ }\bibfield  {title} {\bibinfo
  {title} {{Simplicial SIRS epidemic models with nonlinear incidence rates}},\
  }\href {https://doi.org/10.1063/5.0040518} {\bibfield  {journal} {\bibinfo
  {journal} {Chaos An Interdiscip. J. Nonlinear Sci.}\ }\textbf {\bibinfo
  {volume} {31}},\ \bibinfo {pages} {053112} (\bibinfo {year}
  {2021})}\BibitemShut {NoStop}%
\bibitem [{\citenamefont {Cohen}\ \emph {et~al.}(2003)\citenamefont {Cohen},
  \citenamefont {Havlin},\ and\ \citenamefont {ben-Avraham}}]{Cohen2003}%
  \BibitemOpen
  \bibfield  {author} {\bibinfo {author} {\bibfnamefont {R.}~\bibnamefont
  {Cohen}}, \bibinfo {author} {\bibfnamefont {S.}~\bibnamefont {Havlin}},\ and\
  \bibinfo {author} {\bibfnamefont {D.}~\bibnamefont {ben-Avraham}},\
  }\bibfield  {title} {\bibinfo {title} {Efficient immunization strategies for
  computer networks and populations},\ }\href
  {https://doi.org/10.1103/PhysRevLett.91.247901} {\bibfield  {journal}
  {\bibinfo  {journal} {Phys. Rev. Lett.}\ }\textbf {\bibinfo {volume} {91}},\
  \bibinfo {pages} {247901} (\bibinfo {year} {2003})}\BibitemShut {NoStop}%
\bibitem [{\citenamefont {Madar}\ \emph {et~al.}(2004)\citenamefont {Madar},
  \citenamefont {Kalisky}, \citenamefont {Cohen}, \citenamefont {ben-Avraham},\
  and\ \citenamefont {Havlin}}]{Madar2004}%
  \BibitemOpen
  \bibfield  {author} {\bibinfo {author} {\bibfnamefont {N.}~\bibnamefont
  {Madar}}, \bibinfo {author} {\bibfnamefont {T.}~\bibnamefont {Kalisky}},
  \bibinfo {author} {\bibfnamefont {R.}~\bibnamefont {Cohen}}, \bibinfo
  {author} {\bibfnamefont {D.}~\bibnamefont {ben-Avraham}},\ and\ \bibinfo
  {author} {\bibfnamefont {S.}~\bibnamefont {Havlin}},\ }\bibfield  {title}
  {\bibinfo {title} {{Immunization and epidemic dynamics in complex
  networks}},\ }\href {https://doi.org/10.1140/epjb/e2004-00119-8} {\bibfield
  {journal} {\bibinfo  {journal} {Eur. Phys. J. B}\ }\textbf {\bibinfo {volume}
  {38}},\ \bibinfo {pages} {269} (\bibinfo {year} {2004})}\BibitemShut
  {NoStop}%
\bibitem [{\citenamefont {Chen}\ \emph {et~al.}(2008)\citenamefont {Chen},
  \citenamefont {Paul}, \citenamefont {Havlin}, \citenamefont {Liljeros},\ and\
  \citenamefont {Stanley}}]{Chen2008}%
  \BibitemOpen
  \bibfield  {author} {\bibinfo {author} {\bibfnamefont {Y.}~\bibnamefont
  {Chen}}, \bibinfo {author} {\bibfnamefont {G.}~\bibnamefont {Paul}}, \bibinfo
  {author} {\bibfnamefont {S.}~\bibnamefont {Havlin}}, \bibinfo {author}
  {\bibfnamefont {F.}~\bibnamefont {Liljeros}},\ and\ \bibinfo {author}
  {\bibfnamefont {H.~E.}\ \bibnamefont {Stanley}},\ }\bibfield  {title}
  {\bibinfo {title} {{Finding a Better Immunization Strategy}},\ }\href
  {https://doi.org/10.1103/PhysRevLett.101.058701} {\bibfield  {journal}
  {\bibinfo  {journal} {Phys. Rev. Lett.}\ }\textbf {\bibinfo {volume} {101}},\
  \bibinfo {pages} {058701} (\bibinfo {year} {2008})}\BibitemShut {NoStop}%
\bibitem [{\citenamefont {Masuda}(2009)}]{Masuda2009}%
  \BibitemOpen
  \bibfield  {author} {\bibinfo {author} {\bibfnamefont {N.}~\bibnamefont
  {Masuda}},\ }\bibfield  {title} {\bibinfo {title} {{Immunization of networks
  with community structure}},\ }\href
  {https://doi.org/10.1088/1367-2630/11/12/123018} {\bibfield  {journal}
  {\bibinfo  {journal} {New J. Phys.}\ }\textbf {\bibinfo {volume} {11}},\
  \bibinfo {pages} {123018} (\bibinfo {year} {2009})}\BibitemShut {NoStop}%
\bibitem [{\citenamefont {Pastor-Satorras}\ and\ \citenamefont
  {Vespignani}(2002)}]{Pastor-Satorras2002}%
  \BibitemOpen
  \bibfield  {author} {\bibinfo {author} {\bibfnamefont {R.}~\bibnamefont
  {Pastor-Satorras}}\ and\ \bibinfo {author} {\bibfnamefont {A.}~\bibnamefont
  {Vespignani}},\ }\bibfield  {title} {\bibinfo {title} {{Immunization of
  complex networks}},\ }\href {https://doi.org/10.1103/PhysRevE.65.036104}
  {\bibfield  {journal} {\bibinfo  {journal} {Phys. Rev. E}\ }\textbf {\bibinfo
  {volume} {65}},\ \bibinfo {pages} {036104} (\bibinfo {year}
  {2002})}\BibitemShut {NoStop}%
\bibitem [{\citenamefont {{Van Mieghem}}\ \emph {et~al.}(2011)\citenamefont
  {{Van Mieghem}}, \citenamefont {Stevanovi{\'{c}}}, \citenamefont {Kuipers},
  \citenamefont {Li}, \citenamefont {van~de Bovenkamp}, \citenamefont {Liu},\
  and\ \citenamefont {Wang}}]{VanMieghem2011}%
  \BibitemOpen
  \bibfield  {author} {\bibinfo {author} {\bibfnamefont {P.}~\bibnamefont {{Van
  Mieghem}}}, \bibinfo {author} {\bibfnamefont {D.}~\bibnamefont
  {Stevanovi{\'{c}}}}, \bibinfo {author} {\bibfnamefont {F.}~\bibnamefont
  {Kuipers}}, \bibinfo {author} {\bibfnamefont {C.}~\bibnamefont {Li}},
  \bibinfo {author} {\bibfnamefont {R.}~\bibnamefont {van~de Bovenkamp}},
  \bibinfo {author} {\bibfnamefont {D.}~\bibnamefont {Liu}},\ and\ \bibinfo
  {author} {\bibfnamefont {H.}~\bibnamefont {Wang}},\ }\bibfield  {title}
  {\bibinfo {title} {{Decreasing the spectral radius of a graph by link
  removals}},\ }\href {https://doi.org/10.1103/PhysRevE.84.016101} {\bibfield
  {journal} {\bibinfo  {journal} {Phys. Rev. E}\ }\textbf {\bibinfo {volume}
  {84}},\ \bibinfo {pages} {016101} (\bibinfo {year} {2011})}\BibitemShut
  {NoStop}%
\bibitem [{\citenamefont {Matamalas}\ \emph {et~al.}(2018)\citenamefont
  {Matamalas}, \citenamefont {Arenas},\ and\ \citenamefont
  {G{\'{o}}mez}}]{Matamalas2018}%
  \BibitemOpen
  \bibfield  {author} {\bibinfo {author} {\bibfnamefont {J.~T.}\ \bibnamefont
  {Matamalas}}, \bibinfo {author} {\bibfnamefont {A.}~\bibnamefont {Arenas}},\
  and\ \bibinfo {author} {\bibfnamefont {S.}~\bibnamefont {G{\'{o}}mez}},\
  }\bibfield  {title} {\bibinfo {title} {{Effective approach to epidemic
  containment using link equations in complex networks}},\ }\href
  {https://doi.org/10.1126/sciadv.aau4212} {\bibfield  {journal} {\bibinfo
  {journal} {Sci. Adv.}\ }\textbf {\bibinfo {volume} {4}},\ \bibinfo {pages}
  {eaau4212} (\bibinfo {year} {2018})}\BibitemShut {NoStop}%
\bibitem [{\citenamefont {Costa}\ and\ \citenamefont
  {Ferreira}(2020)}]{Costa2020}%
  \BibitemOpen
  \bibfield  {author} {\bibinfo {author} {\bibfnamefont {G.~S.}\ \bibnamefont
  {Costa}}\ and\ \bibinfo {author} {\bibfnamefont {S.~C.}\ \bibnamefont
  {Ferreira}},\ }\bibfield  {title} {\bibinfo {title} {Nonmassive immunization
  to contain spreading on complex networks},\ }\href
  {https://doi.org/10.1103/PhysRevE.101.022311} {\bibfield  {journal} {\bibinfo
   {journal} {Phys. Rev. E}\ }\textbf {\bibinfo {volume} {101}},\ \bibinfo
  {pages} {022311} (\bibinfo {year} {2020})}\BibitemShut {NoStop}%
\bibitem [{\citenamefont {Shim}(2021)}]{Shim2021}%
  \BibitemOpen
  \bibfield  {author} {\bibinfo {author} {\bibfnamefont {E.}~\bibnamefont
  {Shim}},\ }\bibfield  {title} {\bibinfo {title} {{Optimal Allocation of the
  Limited COVID-19 Vaccine Supply in South Korea}},\ }\href
  {https://doi.org/10.3390/jcm10040591} {\bibfield  {journal} {\bibinfo
  {journal} {J. Clin. Med.}\ }\textbf {\bibinfo {volume} {10}},\ \bibinfo
  {pages} {591} (\bibinfo {year} {2021})}\BibitemShut {NoStop}%
\bibitem [{\citenamefont {{Yang Wang}}\ \emph {et~al.}(2003)\citenamefont
  {{Yang Wang}}, \citenamefont {Chakrabarti}, \citenamefont {{Chenxi Wang}},\
  and\ \citenamefont {Faloutsos}}]{Wang2003}%
  \BibitemOpen
  \bibfield  {author} {\bibinfo {author} {\bibnamefont {{Yang Wang}}}, \bibinfo
  {author} {\bibfnamefont {D.}~\bibnamefont {Chakrabarti}}, \bibinfo {author}
  {\bibnamefont {{Chenxi Wang}}},\ and\ \bibinfo {author} {\bibfnamefont
  {C.}~\bibnamefont {Faloutsos}},\ }\bibfield  {title} {\bibinfo {title}
  {{Epidemic spreading in real networks: an eigenvalue viewpoint}},\ }in\ \href
  {https://doi.org/10.1109/RELDIS.2003.1238052} {\emph {\bibinfo {booktitle}
  {22nd Int. Symp. Reliab. Distrib. Syst. 2003. Proceedings.}}}\ (\bibinfo
  {publisher} {IEEE Comput. Soc},\ \bibinfo {year} {2003})\ pp.\ \bibinfo
  {pages} {25--34}\BibitemShut {NoStop}%
\bibitem [{\citenamefont {G{\'{o}}mez}\ \emph {et~al.}(2010)\citenamefont
  {G{\'{o}}mez}, \citenamefont {Arenas}, \citenamefont {Borge-Holthoefer},
  \citenamefont {Meloni},\ and\ \citenamefont {Moreno}}]{Gomez2010}%
  \BibitemOpen
  \bibfield  {author} {\bibinfo {author} {\bibfnamefont {S.}~\bibnamefont
  {G{\'{o}}mez}}, \bibinfo {author} {\bibfnamefont {A.}~\bibnamefont {Arenas}},
  \bibinfo {author} {\bibfnamefont {J.}~\bibnamefont {Borge-Holthoefer}},
  \bibinfo {author} {\bibfnamefont {S.}~\bibnamefont {Meloni}},\ and\ \bibinfo
  {author} {\bibfnamefont {Y.}~\bibnamefont {Moreno}},\ }\bibfield  {title}
  {\bibinfo {title} {{Discrete-time Markov chain approach to contact-based
  disease spreading in complex networks}},\ }\href
  {https://doi.org/10.1209/0295-5075/89/38009} {\bibfield  {journal} {\bibinfo
  {journal} {EPL}\ }\textbf {\bibinfo {volume} {89}},\ \bibinfo {pages} {38009}
  (\bibinfo {year} {2010})}\BibitemShut {NoStop}%
\bibitem [{\citenamefont {Pastor-Satorras}\ and\ \citenamefont
  {Vespignani}(2001{\natexlab{b}})}]{Pastor-Satorras2001d}%
  \BibitemOpen
  \bibfield  {author} {\bibinfo {author} {\bibfnamefont {R.}~\bibnamefont
  {Pastor-Satorras}}\ and\ \bibinfo {author} {\bibfnamefont {A.}~\bibnamefont
  {Vespignani}},\ }\bibfield  {title} {\bibinfo {title} {{Epidemic dynamics and
  endemic states in complex networks}},\ }\href
  {https://doi.org/10.1103/PhysRevE.63.066117} {\bibfield  {journal} {\bibinfo
  {journal} {Phys. Rev. E}\ }\textbf {\bibinfo {volume} {63}},\ \bibinfo
  {pages} {066117} (\bibinfo {year} {2001}{\natexlab{b}})}\BibitemShut
  {NoStop}%
\bibitem [{\citenamefont {Valdano}\ \emph {et~al.}(2015)\citenamefont
  {Valdano}, \citenamefont {Ferreri}, \citenamefont {Poletto},\ and\
  \citenamefont {Colizza}}]{Valdano2015}%
  \BibitemOpen
  \bibfield  {author} {\bibinfo {author} {\bibfnamefont {E.}~\bibnamefont
  {Valdano}}, \bibinfo {author} {\bibfnamefont {L.}~\bibnamefont {Ferreri}},
  \bibinfo {author} {\bibfnamefont {C.}~\bibnamefont {Poletto}},\ and\ \bibinfo
  {author} {\bibfnamefont {V.}~\bibnamefont {Colizza}},\ }\bibfield  {title}
  {\bibinfo {title} {{Analytical Computation of the Epidemic Threshold on
  Temporal Networks}},\ }\href {https://doi.org/10.1103/PhysRevX.5.021005}
  {\bibfield  {journal} {\bibinfo  {journal} {Phys. Rev. X}\ }\textbf {\bibinfo
  {volume} {5}},\ \bibinfo {pages} {021005} (\bibinfo {year}
  {2015})}\BibitemShut {NoStop}%
\bibitem [{\citenamefont {Lionberger}\ \emph {et~al.}(1968)\citenamefont
  {Lionberger}, \citenamefont {Coleman}, \citenamefont {Katz},\ and\
  \citenamefont {Menzel}}]{Lionberger1968}%
  \BibitemOpen
  \bibfield  {author} {\bibinfo {author} {\bibfnamefont {H.~F.}\ \bibnamefont
  {Lionberger}}, \bibinfo {author} {\bibfnamefont {J.~S.}\ \bibnamefont
  {Coleman}}, \bibinfo {author} {\bibfnamefont {E.}~\bibnamefont {Katz}},\ and\
  \bibinfo {author} {\bibfnamefont {H.}~\bibnamefont {Menzel}},\ }\href
  {https://doi.org/10.2307/2948322} {\emph {\bibinfo {title} {J. Health Soc.
  Behav.}}},\ Vol.~\bibinfo {volume} {9}\ (\bibinfo  {publisher} {Bobbs-Merrill
  Co},\ \bibinfo {year} {1968})\ p.~\bibinfo {pages} {92}\BibitemShut {NoStop}%
\bibitem [{\citenamefont {Heath}\ \emph {et~al.}(2001)\citenamefont {Heath},
  \citenamefont {Bell},\ and\ \citenamefont {Sternberg}}]{Heath2001}%
  \BibitemOpen
  \bibfield  {author} {\bibinfo {author} {\bibfnamefont {C.}~\bibnamefont
  {Heath}}, \bibinfo {author} {\bibfnamefont {C.}~\bibnamefont {Bell}},\ and\
  \bibinfo {author} {\bibfnamefont {E.}~\bibnamefont {Sternberg}},\ }\href
  {https://doi.org/10.1037/0022-3514.81.6.1028} {\bibfield  {journal} {\bibinfo
   {journal} {J. Pers. Soc. Psychol.}\ }\textbf {\bibinfo {volume} {81}},\
  \bibinfo {pages} {1028} (\bibinfo {year} {2001})}\BibitemShut {NoStop}%
\bibitem [{\citenamefont {MacDonald}\ and\ \citenamefont
  {MacDonald}(1964)}]{MacDonald1964}%
  \BibitemOpen
  \bibfield  {author} {\bibinfo {author} {\bibfnamefont {J.~S.}\ \bibnamefont
  {MacDonald}}\ and\ \bibinfo {author} {\bibfnamefont {L.~D.}\ \bibnamefont
  {MacDonald}},\ }\href {https://doi.org/10.2307/3348581} {\bibfield  {journal}
  {\bibinfo  {journal} {Milbank Mem. Fund Q.}\ }\textbf {\bibinfo {volume}
  {42}},\ \bibinfo {pages} {82} (\bibinfo {year} {1964})}\BibitemShut {NoStop}%
\bibitem [{\citenamefont {Crane}(1999)}]{Crane1999}%
  \BibitemOpen
  \bibfield  {author} {\bibinfo {author} {\bibfnamefont {D.}~\bibnamefont
  {Crane}},\ }\href {https://doi.org/10.1177/000271629956600102} {\bibfield
  {journal} {\bibinfo  {journal} {Ann. Am. Acad. Pol. Soc. Sci.}\ }\textbf
  {\bibinfo {volume} {566}},\ \bibinfo {pages} {13} (\bibinfo {year}
  {1999})}\BibitemShut {NoStop}%
\bibitem [{\citenamefont {Goh}\ \emph {et~al.}(2001)\citenamefont {Goh},
  \citenamefont {Kahng},\ and\ \citenamefont {Kim}}]{Goh2001c}%
  \BibitemOpen
  \bibfield  {author} {\bibinfo {author} {\bibfnamefont {K.-I.}\ \bibnamefont
  {Goh}}, \bibinfo {author} {\bibfnamefont {B.}~\bibnamefont {Kahng}},\ and\
  \bibinfo {author} {\bibfnamefont {D.}~\bibnamefont {Kim}},\ }\bibfield
  {title} {\bibinfo {title} {{Universal Behavior of Load Distribution in
  Scale-Free Networks}},\ }\href
  {https://doi.org/10.1103/PhysRevLett.87.278701} {\bibfield  {journal}
  {\bibinfo  {journal} {Phys. Rev. Lett.}\ }\textbf {\bibinfo {volume} {87}},\
  \bibinfo {pages} {278701} (\bibinfo {year} {2001})}\BibitemShut {NoStop}%
\bibitem [{\citenamefont {Lee}\ \emph {et~al.}(2006)\citenamefont {Lee},
  \citenamefont {Goh}, \citenamefont {Kahng},\ and\ \citenamefont
  {Kim}}]{Lee2006}%
  \BibitemOpen
  \bibfield  {author} {\bibinfo {author} {\bibfnamefont {J.-S.}\ \bibnamefont
  {Lee}}, \bibinfo {author} {\bibfnamefont {K.-I.}\ \bibnamefont {Goh}},
  \bibinfo {author} {\bibfnamefont {B.}~\bibnamefont {Kahng}},\ and\ \bibinfo
  {author} {\bibfnamefont {D.}~\bibnamefont {Kim}},\ }\bibfield  {title}
  {{\selectlanguage {English}\bibinfo {title} {{Intrinsic degree-correlations
  in the static model of scale-free networks}}},\ }\href
  {https://doi.org/10.1140/epjb/e2006-00051-y} {\bibfield  {journal} {\bibinfo
  {journal} {Eur. Phys. J. B}\ }\textbf {\bibinfo {volume} {49}},\ \bibinfo
  {pages} {231} (\bibinfo {year} {2006})}\BibitemShut {NoStop}%
\bibitem [{\citenamefont {Goh}\ \emph {et~al.}(2006)\citenamefont {Goh},
  \citenamefont {Salvi}, \citenamefont {Kahng},\ and\ \citenamefont
  {Kim}}]{Goh2006}%
  \BibitemOpen
  \bibfield  {author} {\bibinfo {author} {\bibfnamefont {K.-I.}\ \bibnamefont
  {Goh}}, \bibinfo {author} {\bibfnamefont {G.}~\bibnamefont {Salvi}}, \bibinfo
  {author} {\bibfnamefont {B.}~\bibnamefont {Kahng}},\ and\ \bibinfo {author}
  {\bibfnamefont {D.}~\bibnamefont {Kim}},\ }\bibfield  {title} {\bibinfo
  {title} {{Skeleton and Fractal Scaling in Complex Networks}},\ }\href
  {https://doi.org/10.1103/PhysRevLett.96.018701} {\bibfield  {journal}
  {\bibinfo  {journal} {Phys. Rev. Lett.}\ }\textbf {\bibinfo {volume} {96}},\
  \bibinfo {pages} {018701} (\bibinfo {year} {2006})}\BibitemShut {NoStop}%
\bibitem [{\citenamefont {Yook}\ and\ \citenamefont {Kim}(2018)}]{Yook2018}%
  \BibitemOpen
  \bibfield  {author} {\bibinfo {author} {\bibfnamefont {S.-H.}\ \bibnamefont
  {Yook}}\ and\ \bibinfo {author} {\bibfnamefont {Y.}~\bibnamefont {Kim}},\
  }\bibfield  {title} {\bibinfo {title} {{Two order parameters for the Kuramoto
  model on complex networks}},\ }\href
  {https://doi.org/10.1103/PhysRevE.97.042317} {\bibfield  {journal} {\bibinfo
  {journal} {Phys. Rev. E}\ }\textbf {\bibinfo {volume} {97}},\ \bibinfo
  {pages} {042317} (\bibinfo {year} {2018})}\BibitemShut {NoStop}%
\bibitem [{\citenamefont {Lee}\ \emph {et~al.}(2018)\citenamefont {Lee},
  \citenamefont {Kahng}, \citenamefont {Cho}, \citenamefont {Goh},\ and\
  \citenamefont {Lee}}]{Lee2018}%
  \BibitemOpen
  \bibfield  {author} {\bibinfo {author} {\bibfnamefont {D.}~\bibnamefont
  {Lee}}, \bibinfo {author} {\bibfnamefont {B.}~\bibnamefont {Kahng}}, \bibinfo
  {author} {\bibfnamefont {Y.~S.}\ \bibnamefont {Cho}}, \bibinfo {author}
  {\bibfnamefont {K.-I.}\ \bibnamefont {Goh}},\ and\ \bibinfo {author}
  {\bibfnamefont {D.-S.}\ \bibnamefont {Lee}},\ }\bibfield  {title} {\bibinfo
  {title} {{Recent Advances of Percolation Theory in Complex Networks}},\
  }\href {https://doi.org/10.3938/jkps.73.152} {\bibfield  {journal} {\bibinfo
  {journal} {J. Korean Phys. Soc.}\ }\textbf {\bibinfo {volume} {73}},\
  \bibinfo {pages} {152} (\bibinfo {year} {2018})}\BibitemShut {NoStop}%
\bibitem [{\citenamefont {Kim}\ \emph {et~al.}(2017)\citenamefont {Kim},
  \citenamefont {Park},\ and\ \citenamefont {Kahng}}]{Kim2017}%
  \BibitemOpen
  \bibfield  {author} {\bibinfo {author} {\bibfnamefont {D.-H.}\ \bibnamefont
  {Kim}}, \bibinfo {author} {\bibfnamefont {J.}~\bibnamefont {Park}},\ and\
  \bibinfo {author} {\bibfnamefont {B.}~\bibnamefont {Kahng}},\ }\bibfield
  {title} {\bibinfo {title} {{Enhanced storage capacity with errors in
  scale-free Hopfield neural networks: An analytical study}},\ }\href
  {https://doi.org/10.1371/journal.pone.0184683} {\bibfield  {journal}
  {\bibinfo  {journal} {PLOS ONE}\ }\textbf {\bibinfo {volume} {12}},\ \bibinfo
  {pages} {e0184683} (\bibinfo {year} {2017})}\BibitemShut {NoStop}%
\bibitem [{\citenamefont {Papadopoulos}\ \emph {et~al.}(2015)\citenamefont
  {Papadopoulos}, \citenamefont {Psomas},\ and\ \citenamefont
  {Krioukov}}]{Papadopoulos2015}%
  \BibitemOpen
  \bibfield  {author} {\bibinfo {author} {\bibfnamefont {F.}~\bibnamefont
  {Papadopoulos}}, \bibinfo {author} {\bibfnamefont {C.}~\bibnamefont
  {Psomas}},\ and\ \bibinfo {author} {\bibfnamefont {D.}~\bibnamefont
  {Krioukov}},\ }\bibfield  {title} {\bibinfo {title} {{Network Mapping by
  Replaying Hyperbolic Growth}},\ }\href
  {https://doi.org/10.1109/TNET.2013.2294052} {\bibfield  {journal} {\bibinfo
  {journal} {IEEE/ACM Trans. Netw.}\ }\textbf {\bibinfo {volume} {23}},\
  \bibinfo {pages} {198} (\bibinfo {year} {2015})}\BibitemShut {NoStop}%
\bibitem [{\citenamefont {Perozzi}\ \emph {et~al.}(2014)\citenamefont
  {Perozzi}, \citenamefont {Al-Rfou},\ and\ \citenamefont
  {Skiena}}]{Perozzi2014}%
  \BibitemOpen
  \bibfield  {author} {\bibinfo {author} {\bibfnamefont {B.}~\bibnamefont
  {Perozzi}}, \bibinfo {author} {\bibfnamefont {R.}~\bibnamefont {Al-Rfou}},\
  and\ \bibinfo {author} {\bibfnamefont {S.}~\bibnamefont {Skiena}},\
  }\bibfield  {title} {\bibinfo {title} {{DeepWalk}},\ }in\ \href
  {https://doi.org/10.1145/2623330.2623732} {\emph {\bibinfo {booktitle} {Proc.
  20th ACM SIGKDD Int. Conf. Knowl. Discov. data Min.}}}\ (\bibinfo
  {publisher} {ACM},\ \bibinfo {address} {New York, NY, USA},\ \bibinfo {year}
  {2014})\ pp.\ \bibinfo {pages} {701--710}\BibitemShut {NoStop}%
\bibitem [{\citenamefont {Grover}\ and\ \citenamefont
  {Leskovec}(2016)}]{Grover2016}%
  \BibitemOpen
  \bibfield  {author} {\bibinfo {author} {\bibfnamefont {A.}~\bibnamefont
  {Grover}}\ and\ \bibinfo {author} {\bibfnamefont {J.}~\bibnamefont
  {Leskovec}},\ }\bibfield  {title} {\bibinfo {title} {Node2vec: Scalable
  feature learning for networks},\ }in\ \href
  {https://doi.org/10.1145/2939672.2939754} {\emph {\bibinfo {booktitle} {Proc.
  22nd ACM SIGKDD Int. Conf. Knowl. Discov. Data Min.}}},\ \bibinfo {series and
  number} {KDD '16}\ (\bibinfo  {publisher} {Association for Computing
  Machinery},\ \bibinfo {address} {New York, NY, USA},\ \bibinfo {year}
  {2016})\ p.\ \bibinfo {pages} {855–864}\BibitemShut {NoStop}%
\bibitem [{\citenamefont {Estrada}\ and\ \citenamefont
  {Rodr{\'{i}}guez-Vel{\'{a}}zquez}(2006)}]{Estrada2005}%
  \BibitemOpen
  \bibfield  {author} {\bibinfo {author} {\bibfnamefont {E.}~\bibnamefont
  {Estrada}}\ and\ \bibinfo {author} {\bibfnamefont {J.~A.}\ \bibnamefont
  {Rodr{\'{i}}guez-Vel{\'{a}}zquez}},\ }\bibfield  {title} {\bibinfo {title}
  {{Subgraph centrality and clustering in complex hyper-networks}},\ }\href
  {https://doi.org/10.1016/j.physa.2005.12.002} {\bibfield  {journal} {\bibinfo
   {journal} {Physica A}\ }\textbf {\bibinfo {volume} {364}},\ \bibinfo {pages}
  {581} (\bibinfo {year} {2006})}\BibitemShut {NoStop}%
\bibitem [{\citenamefont {Harary}\ and\ \citenamefont
  {Kommel}(1979)}]{Harary1979}%
  \BibitemOpen
  \bibfield  {author} {\bibinfo {author} {\bibfnamefont {F.}~\bibnamefont
  {Harary}}\ and\ \bibinfo {author} {\bibfnamefont {H.~J.}\ \bibnamefont
  {Kommel}},\ }\bibfield  {title} {\bibinfo {title} {{Matrix measures for
  transitivity and balance*}},\ }\href
  {https://doi.org/10.1080/0022250X.1979.9989889} {\bibfield  {journal}
  {\bibinfo  {journal} {J. Math. Sociol.}\ }\textbf {\bibinfo {volume} {6}},\
  \bibinfo {pages} {199} (\bibinfo {year} {1979})}\BibitemShut {NoStop}%
\bibitem [{\citenamefont {Papadopoulos}\ \emph {et~al.}(2012)\citenamefont
  {Papadopoulos}, \citenamefont {Kitsak}, \citenamefont {Serrano},
  \citenamefont {Bogu{\~{n}}{\'{a}}},\ and\ \citenamefont
  {Krioukov}}]{Papadopoulos2012}%
  \BibitemOpen
  \bibfield  {author} {\bibinfo {author} {\bibfnamefont {F.}~\bibnamefont
  {Papadopoulos}}, \bibinfo {author} {\bibfnamefont {M.}~\bibnamefont
  {Kitsak}}, \bibinfo {author} {\bibfnamefont {M.~{\'{A}}.}\ \bibnamefont
  {Serrano}}, \bibinfo {author} {\bibfnamefont {M.}~\bibnamefont
  {Bogu{\~{n}}{\'{a}}}},\ and\ \bibinfo {author} {\bibfnamefont
  {D.}~\bibnamefont {Krioukov}},\ }\bibfield  {title} {\bibinfo {title}
  {{Popularity versus similarity in growing networks}},\ }\href
  {https://doi.org/10.1038/nature11459} {\bibfield  {journal} {\bibinfo
  {journal} {Nature}\ }\textbf {\bibinfo {volume} {489}},\ \bibinfo {pages}
  {537} (\bibinfo {year} {2012})}\BibitemShut {NoStop}%
\bibitem [{\citenamefont {P{\'{e}}rez-Espigares}\ \emph
  {et~al.}(2017)\citenamefont {P{\'{e}}rez-Espigares}, \citenamefont
  {Marcuzzi}, \citenamefont {Guti{\'{e}}rrez},\ and\ \citenamefont
  {Lesanovsky}}]{Perez-Espigares2017a}%
  \BibitemOpen
  \bibfield  {author} {\bibinfo {author} {\bibfnamefont {C.}~\bibnamefont
  {P{\'{e}}rez-Espigares}}, \bibinfo {author} {\bibfnamefont {M.}~\bibnamefont
  {Marcuzzi}}, \bibinfo {author} {\bibfnamefont {R.}~\bibnamefont
  {Guti{\'{e}}rrez}},\ and\ \bibinfo {author} {\bibfnamefont {I.}~\bibnamefont
  {Lesanovsky}},\ }\bibfield  {title} {\bibinfo {title} {{Epidemic Dynamics in
  Open Quantum Spin Systems}},\ }\href
  {https://doi.org/10.1103/PhysRevLett.119.140401} {\bibfield  {journal}
  {\bibinfo  {journal} {Phys. Rev. Lett.}\ }\textbf {\bibinfo {volume} {119}},\
  \bibinfo {pages} {140401} (\bibinfo {year} {2017})}\BibitemShut {NoStop}%
\bibitem [{\citenamefont {Lim}(2005)}]{Lim2005}%
  \BibitemOpen
  \bibfield  {author} {\bibinfo {author} {\bibfnamefont {L.-H.}\ \bibnamefont
  {Lim}},\ }\bibfield  {title} {\bibinfo {title} {Singular values and
  eigenvalues of tensors: a variational approach},\ }in\ \href
  {https://doi.org/10.1109/CAMAP.2005.1574201} {\emph {\bibinfo {booktitle}
  {1st IEEE Int. Work. Comput. Adv. Multi-Sensor Adapt. Process. 2005.}}}\
  (\bibinfo {year} {2005})\ pp.\ \bibinfo {pages} {129--132}\BibitemShut
  {NoStop}%
\bibitem [{\citenamefont {Qi}(2005)}]{Qi2005}%
  \BibitemOpen
  \bibfield  {author} {\bibinfo {author} {\bibfnamefont {L.}~\bibnamefont
  {Qi}},\ }\bibfield  {title} {\bibinfo {title} {{Eigenvalues of a real
  supersymmetric tensor}},\ }\href {https://doi.org/10.1016/j.jsc.2005.05.007}
  {\bibfield  {journal} {\bibinfo  {journal} {J. Symb. Comput.}\ }\textbf
  {\bibinfo {volume} {40}},\ \bibinfo {pages} {1302} (\bibinfo {year}
  {2005})}\BibitemShut {NoStop}%
\bibitem [{\citenamefont {Comon}\ \emph {et~al.}(2008)\citenamefont {Comon},
  \citenamefont {Golub}, \citenamefont {Lim},\ and\ \citenamefont
  {Mourrain}}]{Comon2008}%
  \BibitemOpen
  \bibfield  {author} {\bibinfo {author} {\bibfnamefont {P.}~\bibnamefont
  {Comon}}, \bibinfo {author} {\bibfnamefont {G.}~\bibnamefont {Golub}},
  \bibinfo {author} {\bibfnamefont {L.-H.}\ \bibnamefont {Lim}},\ and\ \bibinfo
  {author} {\bibfnamefont {B.}~\bibnamefont {Mourrain}},\ }\bibfield  {title}
  {\bibinfo {title} {{Symmetric Tensors and Symmetric Tensor Rank}},\ }\href
  {https://doi.org/10.1137/060661569} {\bibfield  {journal} {\bibinfo
  {journal} {SIAM J. Matrix Anal. Appl.}\ }\textbf {\bibinfo {volume} {30}},\
  \bibinfo {pages} {1254} (\bibinfo {year} {2008})}\BibitemShut {NoStop}%
\bibitem [{\citenamefont {Ferreira}\ \emph {et~al.}(2011)\citenamefont
  {Ferreira}, \citenamefont {Ferreira},\ and\ \citenamefont
  {Pastor-Satorras}}]{Ferreira2011}%
  \BibitemOpen
  \bibfield  {author} {\bibinfo {author} {\bibfnamefont {S.~C.}\ \bibnamefont
  {Ferreira}}, \bibinfo {author} {\bibfnamefont {R.~S.}\ \bibnamefont
  {Ferreira}},\ and\ \bibinfo {author} {\bibfnamefont {R.}~\bibnamefont
  {Pastor-Satorras}},\ }\bibfield  {title} {\bibinfo {title} {{Quasistationary
  analysis of the contact process on annealed scale-free networks}},\ }\href
  {https://doi.org/10.1103/PhysRevE.83.066113} {\bibfield  {journal} {\bibinfo
  {journal} {Phys. Rev. E}\ }\textbf {\bibinfo {volume} {83}},\ \bibinfo
  {pages} {066113} (\bibinfo {year} {2011})}\BibitemShut {NoStop}%
\bibitem [{\citenamefont {Ferreira}\ \emph {et~al.}(2012)\citenamefont
  {Ferreira}, \citenamefont {Castellano},\ and\ \citenamefont
  {Pastor-Satorras}}]{Ferreira2012}%
  \BibitemOpen
  \bibfield  {author} {\bibinfo {author} {\bibfnamefont {S.~C.}\ \bibnamefont
  {Ferreira}}, \bibinfo {author} {\bibfnamefont {C.}~\bibnamefont
  {Castellano}},\ and\ \bibinfo {author} {\bibfnamefont {R.}~\bibnamefont
  {Pastor-Satorras}},\ }\bibfield  {title} {\bibinfo {title} {{Epidemic
  thresholds of the susceptible-infected-susceptible model on networks: A
  comparison of numerical and theoretical results}},\ }\href
  {https://doi.org/10.1103/PhysRevE.86.041125} {\bibfield  {journal} {\bibinfo
  {journal} {Phys. Rev. E}\ }\textbf {\bibinfo {volume} {86}},\ \bibinfo
  {pages} {041125} (\bibinfo {year} {2012})}\BibitemShut {NoStop}%
\bibitem [{\citenamefont {Holme}\ and\ \citenamefont {Kim}(2002)}]{Holme2002}%
  \BibitemOpen
  \bibfield  {author} {\bibinfo {author} {\bibfnamefont {P.}~\bibnamefont
  {Holme}}\ and\ \bibinfo {author} {\bibfnamefont {B.~J.}\ \bibnamefont
  {Kim}},\ }\bibfield  {title} {\bibinfo {title} {{Growing scale-free networks
  with tunable clustering}},\ }\href
  {https://doi.org/10.1103/PhysRevE.65.026107} {\bibfield  {journal} {\bibinfo
  {journal} {Phys. Rev. E}\ }\textbf {\bibinfo {volume} {65}},\ \bibinfo
  {pages} {026107} (\bibinfo {year} {2002})}\BibitemShut {NoStop}%
\bibitem [{\citenamefont {Jo}\ and\ \citenamefont {Kahng}(2020)}]{Jo2020}%
  \BibitemOpen
  \bibfield  {author} {\bibinfo {author} {\bibfnamefont {M.}~\bibnamefont
  {Jo}}\ and\ \bibinfo {author} {\bibfnamefont {B.}~\bibnamefont {Kahng}},\
  }\bibfield  {title} {\bibinfo {title} {{Tricritical directed percolation with
  long-range interaction in one and two dimensions}},\ }\href
  {https://doi.org/10.1103/PhysRevE.101.022121} {\bibfield  {journal} {\bibinfo
   {journal} {Phys. Rev. E}\ }\textbf {\bibinfo {volume} {101}},\ \bibinfo
  {pages} {022121} (\bibinfo {year} {2020})}\BibitemShut {NoStop}%
\bibitem [{\citenamefont {Im}\ and\ \citenamefont {Kahng}(2018)}]{Im2018}%
  \BibitemOpen
  \bibfield  {author} {\bibinfo {author} {\bibfnamefont {Y.~S.}\ \bibnamefont
  {Im}}\ and\ \bibinfo {author} {\bibfnamefont {B.}~\bibnamefont {Kahng}},\
  }\bibfield  {title} {\bibinfo {title} {{Dismantling efficiency and network
  fractality}},\ }\href {https://doi.org/10.1103/PhysRevE.98.012316} {\bibfield
   {journal} {\bibinfo  {journal} {Phys. Rev. E}\ }\textbf {\bibinfo {volume}
  {98}},\ \bibinfo {pages} {012316} (\bibinfo {year} {2018})}\BibitemShut
  {NoStop}%
\bibitem [{\citenamefont {St-Onge}\ \emph
  {et~al.}(2021{\natexlab{b}})\citenamefont {St-Onge}, \citenamefont
  {Thibeault}, \citenamefont {Allard}, \citenamefont {Dub{\'{e}}},\ and\
  \citenamefont {H{\'{e}}bert-Dufresne}}]{St-Onge2021a}%
  \BibitemOpen
  \bibfield  {author} {\bibinfo {author} {\bibfnamefont {G.}~\bibnamefont
  {St-Onge}}, \bibinfo {author} {\bibfnamefont {V.}~\bibnamefont {Thibeault}},
  \bibinfo {author} {\bibfnamefont {A.}~\bibnamefont {Allard}}, \bibinfo
  {author} {\bibfnamefont {L.~J.}\ \bibnamefont {Dub{\'{e}}}},\ and\ \bibinfo
  {author} {\bibfnamefont {L.}~\bibnamefont {H{\'{e}}bert-Dufresne}},\
  }\bibfield  {title} {\bibinfo {title} {{Social Confinement and Mesoscopic
  Localization of Epidemics on Networks}},\ }\href
  {https://doi.org/10.1103/PhysRevLett.126.098301} {\bibfield  {journal}
  {\bibinfo  {journal} {Phys. Rev. Lett.}\ }\textbf {\bibinfo {volume} {126}},\
  \bibinfo {pages} {098301} (\bibinfo {year} {2021}{\natexlab{b}})}\BibitemShut
  {NoStop}%
\bibitem [{\citenamefont {Burgio}\ \emph {et~al.}(2021)\citenamefont {Burgio},
  \citenamefont {Arenas}, \citenamefont {G{\'{o}}mez},\ and\ \citenamefont
  {Matamalas}}]{Burgio2021}%
  \BibitemOpen
  \bibfield  {author} {\bibinfo {author} {\bibfnamefont {G.}~\bibnamefont
  {Burgio}}, \bibinfo {author} {\bibfnamefont {A.}~\bibnamefont {Arenas}},
  \bibinfo {author} {\bibfnamefont {S.}~\bibnamefont {G{\'{o}}mez}},\ and\
  \bibinfo {author} {\bibfnamefont {J.~T.}\ \bibnamefont {Matamalas}},\
  }\bibfield  {title} {\bibinfo {title} {{Network clique cover approximation to
  analyze complex contagions through group interactions}},\ }\href
  {https://doi.org/10.1038/s42005-021-00618-z} {\bibfield  {journal} {\bibinfo
  {journal} {Commun. Phys.}\ }\textbf {\bibinfo {volume} {4}},\ \bibinfo
  {pages} {111} (\bibinfo {year} {2021})}\BibitemShut {NoStop}%


\end{thebibliography}
    %

\end{document}